\documentclass[aps,prd,onecolumn,preprintnumbers,groupedaddress,showpacs,nofootinbib,amssymb]{revtex4-2}
\usepackage{graphicx,color}
\usepackage{amsmath}
\usepackage{amssymb}
\usepackage{amsfonts}
\usepackage{bm}
\usepackage{cancel}
\usepackage{slashed}
\usepackage{hyperref}

\usepackage{tikz}

\newcommand{\e}{\mathrm{e}}

\allowdisplaybreaks[4]
\begin{document}

\preprint{KEK-TH-2852, KEK-Cosmo-0425}

\title{ACT-DR6 consistent inflation in generalised entropic cosmology and $f(Q)$ gravity}

\author{Shin'ichi~Nojiri$^{1,2,3}$}
\email{nojiri@nagoya-u.jp}

\author{Sergei~D.~Odintsov$^{4,5}$}
\email{odintsov@ice.csic.es} 

\affiliation{$^{1)}$ Tsung-Dao Lee Institute (TDLI), Shanghai Jiao Tong University, No. 1 Lisuo Road, Pudong New Area, Shanghai, 201210, China}

\affiliation{$^{2)}$ KEK Theory Center, Institute of Particle and Nuclear Studies, High Energy Accelerator Research Organization (KEK), Oho 1-1, Tsukuba, Ibaraki 305-0801, Japan}

\affiliation{$^{3)}$ Kobayashi-Maskawa Institute for the Origin of Particles and the Universe, Nagoya University, Nagoya 464-8602, Japan}

\affiliation{$^{4)}$ Institute of Space Sciences,
(ICE/CSIC-IEEC), Carrer de Can Magrans s/n, 08193 Barcelona, Spain}

\affiliation{$^{5)}$ Instituci\'o Catalana de Recerca i Estudis Avan\c{c}ats (ICREA),
Passeig Luis Companys, 23, 08010 Barcelona, Spain}

\begin{abstract}
Within the framework of generalised entropic cosmology and its equivalent $f(Q)$ gravity, we construct inflationary scenarios consistent with recent Atacama Cosmology Telescope (ACT) observations combined with Planck-BAO data. 
We work with $f(Q)$ gravity and use the reconstruction scheme from which one can obtain an arbitrary cosmological evolution consistent with desirable theoretical or observational considerations. 
Explicit inflationary cosmologies of the above theory, which pass ACT-Data Release 6 (DR6) data combined with Planck and BAO, are obtained. 
For $f(Q)$ gravity coupled with scalar, we obtain ACT inflation for large classes of $f(Q)$ via the choice of the scalar potentials. 
The entropic cosmology based on generalised entropy, which includes most of the known entropies like Tsallis, R\'{e}nyi, Barrow, etc as a particular case, is investigated. 
The corresponding FRW equations, depending on the parameters of generalised entropy, are presented.
Using the correspondence between the generalised entropy and $f(Q)$ gravity, we obtain generalised entropic inflation, which is consistent with the ACT-DR6. 
At the next step, the entropic gravity coupled with scalar is considered. 
It is demonstrated how one can obtain ACT-DR6-consistent inflation for such a theory based on any generalised entropy considered in this paper via the choice of scalar potentials. 
These findings may point towards the correct unifying entropy of the early universe.
\end{abstract}

\maketitle

\section{Introduction\label{Sec1}}

Recently, the Atacama Cosmology Telescope (ACT) Data Release 6 (DR6) \cite{AtacamaCosmologyTelescope:2025vnj, AtacamaCosmologyTelescope:2025nti, CosmoVerseNetwork:2025alb} renewed some constraints on the parameters of inflation, like the spectral index parameter $n_s$ and the tensor-to-scalar ratio parameter $r$~\cite{AtacamaCosmologyTelescope:2025blo, AtacamaCosmologyTelescope:2025nti}. 
Especially, the median of $n_s$ becomes slightly higher than the value given in the Planck observation~\cite{Planck:2018jri} (see also \cite{Planck:2013jfk, Planck:2015sxf}). 
The shift changes the constraints on the models of inflation. 
After that, many papers tried to verify that the old inflationary models could still be correct or be modified/proposed new models compatible with the ACT constraints (for the mainstream of ACT inflation works, see
\cite{Yang:2026rzn, Gonuguntla:2026rkw, CastroJunior:2026gnm, Odintsov:2026dss, Yang:2026flt, Zambrano:2026fau, Latosh:2026ckf, Yogesh:2026esn, Ahmed:2026agd, Whittingham:2026cbo, Yuennan:2026fcn, Ahmed:2026msg, Peng:2026ofs, Nojiri:2026hij, Chattopadhyay:2026yam, Modak:2025grj, Thakur:2025uua, Wang:2025cpp, McDonough:2025lzo, Keskin:2025zqq, Chakraborty:2025wqn, Yuennan:2025mlg, Bezerra-Sobrinho:2025gfg, Fu:2025ciy, Afshar:2025ndm, Qiu:2025iqm, Yuennan:2025tyx, Pozdeeva:2025wsl, Hell:2025lbl, Zhu:2025twm, Ahmed:2025sfm,Yi:2025dms, Nojiri:2026hij, Odintsov:2025eiv, Kallosh:2025rni, Gao:2025onc, Liu:2025qca, Yogesh:2025wak, Yi:2025dms, Peng:2025bws, Yin:2025rrs, Byrnes:2025kit, Wolf:2025ecy, Aoki:2025wld, Gao:2025viy, Zahoor:2025nuq, Ferreira:2025lrd, Mohammadi:2025gbu, Choudhury:2025vso, Odintsov:2025wai, Q:2025ycf, Zhu:2025twm, Kouniatalis:2025orn, Hai:2025wvs, Dioguardi:2025vci, Yuennan:2025kde, Ajith:2025rvf, Kuralkar:2025hoz, Modak:2025bjv, Aoki:2025ywt, Ahghari:2025hfy, McDonough:2025lzo, Chakraborty:2025wqn, NooriGashti:2025gug, Yuennan:2025mlg, Deb:2025gtk, Afshar:2025ndm, Ellis:2025zrf, Iacconi:2025odq, Yuennan:2025tyx, Wang:2025cpp, Qiu:2025iqm, Wang:2025dbj, Asaka:2015vza, Oikonomou:2025htz, Choudhury:2025hnu, Singh:2025uyr, Kim:2025dyi, Peng:2026ofs, Garny:2026gcs, Odintsov:2026doe, DOnofrio:2025bol, Odintsov:2025eiv, Odintsov:2025jky, Oikonomou:2026qkj, Odintsov:2026cxz, Yang:2026rzn, Yang:2026flt}). 
In particular, in \cite{Nojiri:2026hij}, a ghost-free non-local $F(R)$ gravity theory has been investigated, and in \cite{Odintsov:2025eiv}, a power-law $F(R)$ gravity has been investigated, showing that these models can provide inflation which satisfies the ACT constraints. 

The main purpose of this work is to construct the generalised entropic inflation consistent with ACT-DR6 constraints. 
In order to do so, we first investigate $f(Q)$ gravity, which is based on the non-metricity scalar $Q$. 
Then the equivalence between such a theory and entropic cosmology is used. 
Finally, we explicitly construct models that satisfy the ACT-DR6 constraints. 

The gravity theories with non-metricity have been actively discussed \cite{Nester:1998mp, BeltranJimenez:2018vdo, Runkla:2018xrv, Capozziello:2022tvv}, for review, see \cite{Heisenberg:2023lru}. 
In the non-metricity gravity, the non-metricity scalar $Q$ is a fundamental geometrical quantity. 
The connection is regarded as a variable independent of the metric in such theories. 
In the non-metricity gravity, we impose the conditions that the Riemann tensor and torsion tensor vanish. 
As the solution of the conditions, the connection is written in terms of four scalar fields \cite{Blixt:2023kyr, Adak:2018vzk, Tomonari:2023wcs}. 
We often choose the scalar fields so that the connection vanishes. 
It is then called a coincident gauge. 
As a result, the metric becomes the only dynamical variable. 
Because the difference between $Q$ and the scalar curvature $R$ in Einstein's gravity is a total derivative, the theory where the action is linear in $Q$ is equivalent to Einstein's gravity and is called symmetric teleparallel equivalent to general relativity. 

As an extension of the symmetric teleparallel theory, we may consider $f(Q)$ gravity theory, where the action is defined by a function $f(Q)$ of $Q$. 
This theory is analogous to $f(R)$ gravity or $f(T)$ gravity with a torsion $T$, but the $f(Q)$ gravity is not equivalent to $f(R)$ or $f(T)$ gravity. 
In $f(Q)$ gravity, the problem of the dynamical degrees of freedom (DOF) has been discussed \cite{Hu:2022anq, DAmbrosio:2023asf, Heisenberg:2023lru, Paliathanasis:2023pqp, Dimakis:2021gby, Hu:2023gui}. 
Despite long-standing discussion, the problem has not been completely solved. 
Still, it has been shown that the only propagating mode in the flat background is a graviton, as in Einstein's theory \cite{Capozziello:2024vix}, as in the situation of the $f(T)$ gravity \cite{Bamba:2013ooa}. 
In the case of the $f(R)$ gravity, however, an extra propagating scalar mode appears. 

As proposed in \cite{Nojiri:2024zab}, we define $f(Q)$ gravity theory by using the metric and only four scalar fields. 
Although it is not clear how many dynamical degrees of freedom on-shell exist, the choice of the independent fields makes the equations given by the variation of the action with respect to these fields well-defined. 

In the above formulation, we construct models consistent with the ACT observation by using the formulation of the ``reconstruction''. 
We usually solve the equations for a given gravity model and find the structure of spacetime as the solution. 
The inverse problem is to find a model that realises the geometry desired from the theoretical and/or observational viewpoints. 
Such a systematic formulation is often called the ``reconstruction''. 
If the model can consistently realise the spacetime, the model could be realistic. 
If we cannot find any consistent model in some class of the theory by the reconstruction formulation, we may discard the class. 
For the Friedmann-Lema\^{i}tre-Robertson-Walker (FLRW) spacetime, the consistent formulation of the reconstruction has been well-studied for the scalar-tensor theories \cite{Nojiri:2005pu, Capozziello:2005tf}, the Einstein--scalar--Gauss-Bonnet gravity \cite{Nojiri:2006je, Nojiri:2005jg}, and $F (\mathcal{G})$ gravity \cite{Cognola:2006eg, Nojiri:2019dwl}, where $\mathcal{G}$ is the Gauss-Bonnet topological invariant, and $f(R)$ gravity \cite{Nojiri:2006gh, Nojiri:2009kx}. 
Some attempts for reconstruction in $f(Q)$ gravity have been considered in \cite{Capozziello:2022wgl, Gadbail:2023klq, Gadbail:2023mvu, Kaczmarek:2024yju, Nojiri:2024zab}. 
In this paper, we use the formulation in \cite{Nojiri:2024zab}. 

We also construct inflationary generalised entropic models so that the models become consistent with the ACT constraints. 
The thermodynamics of black holes started from the fact that the black hole has an entropy called Bekenstein-Hawking entropy~\cite{Bekenstein:1973ur, Hawking:1975vcx}. 
By using the Bekenstein-Hawking entropy, the black hole thermodynamics and entropic cosmology related to the apparent horizon have been well studied. 

Besides the standard Gibbs and Boltzmann entropies, depending on different physical and statistical systems, various entropies have been proposed like Tsallis entropy~\cite{Tsallis:1987eu}, R\'{e}nyi entropy~\cite{Renyi}, Barrow entropy~\cite{Barrow:2020tzx}, Sharma-Mittal entropy~\cite{SayahianJahromi:2018irq}, Kaniadakis entropy~\cite{Kaniadakis:2005zk, Drepanou:2021jiv}, Loop Quantum Gravity entropy~\cite{Majhi:2017zao, Liu:2021dvj}, four- and six-parameter generalised entropies~\cite{Nojiri:2022aof, Nojiri:2022dkr}, three-parameter generalised entropy~\cite{Nojiri:2022aof} and five-parameter entropy~\cite{Odintsov:2022qnn}.
Based on the above entropies, modified FRLW equations can be derived. 

On the other hand, Jacobson has shown that the Einstein equation can be derived from the Bekenstein-Hawking entropy by using the thermodynamical consideration~\cite{Jacobson:1995ab}. 
We can also obtain modified FLRW equations for different entropies (entropic cosmology), see \cite{Nojiri:2024zdu} for the review. 
The gravity sector of various entropies could correspond to the modified gravity theories ( see reviews~\cite{Nojiri:2010wj, Capozziello:2011et, Nojiri:2017ncd}), but some points are not clear. 
The key solution could be cosmological equations. 
By applying the Bekenstein-Hawking entropy for a cosmological apparent horizon and by using the thermodynamical considerations, the first Friedmann equation, which is derived from the Einstein equation, has been obtained~\cite{Padmanabhan:2009vy, Cai:2005ra}. 
In \cite{Nojiri:2025fiu}, the correspondence between the generalised entropies and $f(T)$ and $f(Q)$ gravities has been clarified via the first Friedmann equation modified by the generalised entropies. 
In this paper, we construct the entropic inflationary cosmology consistent with the ACT constraints based on the above equivalence. 
In other words, first we derive such consistent inflation in $f(Q)$ gravity. 

In the next section, we briefly review the non-metricity gravity. 
In Section~\ref{Sec3}, we explain how the model realising a given expansion history of the universe can be reconstructed. 
In Section~\ref{Sec4}, via the formulation of the reconstruction in Section~\ref{Sec3}, we propose two models of inflation. 
One model is given by using the cosmological time, and another one is described by the $e$-folding number. 
In Section~\ref{Sec5}, for the models constructed in Section~\ref{Sec4}, we investigate if they can satisfy the constraints given by the ACT-DR6, etc. 
We show that the models could be realistic in this sense. 
In Section~\ref{SecVI}, we consider  $f(Q)$ theory coupled with a scalar field and construct inflation consistent with the ACT-DR6 constraints. 
In Section~\ref{Sec6}, we review and clarify the correspondence between the $f(Q)$ gravity and the generalised entropic cosmology. 
By using the correspondence, we construct an inflationary entropic cosmology that satisfies the ACT-DR6 constraints. Note that very few theories may lead to ACT-consistent inflation.
Eventually, the corresponding generalised entropy may be conjectured to be the entropy of the primordial universe. 
We then investigate generalised entropic gravity coupled with a scalar.
It is explicitly demonstrated how such a theory may give ACT-consistent inflation by the corresponding choice of scalar potential. 
In fact, eventually, a very large class of generalised entropies may lead to ACT inflation via the corresponding choice of scalar potentials.
The last Section is devoted to a summary and discussion.

\section{Brief review of non-metricity gravity\label{Sec2}}

We start with a brief review of non-metricity gravity.

General affine connections are given by 
\begin{align}
\label{affine}
{\Gamma^\sigma}_{\mu \nu}= {{\tilde \Gamma}^\sigma}_{\mu \nu} + K^\sigma_{\;\mu \nu} + L^\sigma_{\;\mu \nu}\,.
\end{align}
Here ${\tilde \Gamma^\sigma}_{\mu \nu}$ is the Levi-Civita connection given by the metric as in general relativity,
\begin{align}
\label{Levi-Civita}
{{\tilde\Gamma}^\sigma}_{\mu \nu} = \frac{1}{2} g^{\sigma \rho} \left( \partial_\mu g_{\rho \nu} + \partial_\nu g_{\rho \mu}- \partial_\rho g_{\mu \nu}\right)\, ,
\end{align}
${K^\sigma}_{\mu \nu}$ is called contortion and defined by using the torsion tensor, ${T^\sigma}_{\mu \nu}={\Gamma^\sigma}_{\mu \nu} - {\Gamma^\sigma}_{\nu \mu}$, as follows, 
\begin{align}
\label{contortion}
{K^\sigma}_{\mu \nu}= \frac{1}{2} \left( {T^\sigma}_{\mu \nu} + T^{\ \sigma}_{\mu\ \nu} + T^{\ \sigma}_{\nu\ \mu} \right) \, ,
\end{align}
${L^\sigma}_{\mu \nu}$ is called deformation and given by 
\begin{align}
\label{deformation}
{L^\sigma}_{\mu \nu}= \frac{1}{2} \left( Q^\sigma_{\;\mu \nu} - Q^{\ \sigma}_{\mu\ \nu} - Q^{\ \sigma}_{\nu\ \mu} \right)\,.
\end{align}
Here ${Q^\sigma}_{\mu \nu}$ is non-metricity tensor defined as, 
\begin{align}
\label{non-metricity}
Q_{\sigma \mu \nu}= \nabla_\sigma g_{\mu \nu}= \partial_\sigma g_{\mu \nu} - {\Gamma^\rho}_{\sigma \mu } g_{\nu \rho} 
 - {\Gamma^\rho}_{\sigma \nu } g_{\mu \rho } \,.
\end{align}
Although we can construct the teleparallel gravity by using the torsion, we can construct another gravity based on non-metricity. 

In both of the teleparallel gravity and non-metricity gravity, we impose the condition that the Riemann tensor constructed from the general affine connections \eqref{affine} vanishes, 
\begin{align}
\label{Riemannzero}
0= R^\lambda_{\ \ \mu\rho\nu}
= -\Gamma^\lambda_{\mu\rho,\nu} + \Gamma^\lambda_{\mu\nu,\rho} - \Gamma^\eta_{\mu\rho}\Gamma^\lambda_{\nu\eta} + \Gamma^\eta_{\mu\nu}\Gamma^\lambda_{\rho\eta} \, . 
\end{align} 
In the teleparallel gravity, we impose one more constraint ${L^\sigma}_{\mu \nu}=0$ and use the torsion scalar $T$ defined as follows, 
\begin{align}
\label{eq:2.4}
T \equiv {S_\rho}^{\mu\nu} {T^\rho}_{\mu\nu}\, , \quad 
{S_\rho}^{\mu\nu} \equiv \frac{1}{2} \left({K^{\mu\nu}}_\rho + {\delta^\mu}_\rho {T^{\alpha\nu}}_{\alpha} - {\delta^\nu}_\rho {T^{\alpha \mu}}_\alpha \right)\, .
\end{align}
As an analogy of the $F(R)$ gravity, we may consider the gravity theory, where the action is given by the torsion scalar $T$ as $f(T)$, which is $f(T)$ gravity. 
There is, however, a serious problem in $f(T)$ gravity. 
It has been shown that the $f(T)$ gravity model includes superluminal propagating modes, which appear non-perturbatively~\cite{Izumi:2012qj, Ong:2013qja}. 
Therefore, the $f(T)$ gravity cannot be a physically consistent theory, and it might be an effective theory of a more fundamental theory. 
Then, in this paper, we concentrate on the study of the gravity theory based on non-metricity. 

The non-metricity gravity is defined by imposing the following conditions: 
There is no torsion, which tells ${\Gamma^\sigma}_{\mu\nu} = {\Gamma^\sigma}_{\nu\mu}$. 
The Riemann curvature vanishes as in \eqref{Riemannzero}. 
The solution of the above conditions is written in terms of four scalar fields $\xi^a$ $\left( a = 0,1,2,3 \right)$, 
\begin{align}
\label{G1B}
{\Gamma^\rho}_{\mu\nu}=\frac{\partial x^\rho}{\partial \xi^a} \partial_\mu \partial_\nu \xi^a \, .
\end{align}
We should note that $\xi^a$'s should be scalar fields and $e^a_\mu\equiv \partial_\mu \xi^a$'s could be identified with vierbein fields. 

The system is invariant under general coordinate transformations. 
Therefore, we can often choose the gauge condition ${\Gamma^\rho}_{\mu\nu}=0$, which is called coincident gauge and realised by choosing $\xi^a$ as follows, 
\begin{align}
\label{Cgauge}
\xi^a=x^a \, . 
\end{align}
The gauge condition, however, often contradicts the standard metric choices of the FLRW universe with a non-flat spatial part and the spherically symmetric spacetime, although we can choose the coincident gauge~\eqref{Cgauge} in the FLRW universe with a flat spatial part \cite{BeltranJimenez:2019tme}. 

Under the infinitesimal transformation, $\xi^a \to \xi^a + \delta \xi^a$, the variation of the connection is given by 
\begin{align}
\label{G1}
\Gamma^\rho_{\mu\nu} \to \Gamma^\rho_{\mu\nu} + \delta \Gamma^\rho_{\mu\nu}
\equiv \Gamma^\rho_{\mu\nu} - \frac{\partial x^\rho}{\partial \xi^a} \partial_\sigma \delta\xi^a\frac{\partial x^\sigma}{\partial \xi^b}\partial_\mu \partial_\nu \xi^b
+ \frac{\partial x^\rho}{\partial \xi^a} \partial_\mu \partial_\nu \delta \xi^a \, .
\end{align}
Because $\xi^a$'s are scalar fields under the infinitesimal coordinate transformation $x^\mu\to x^\mu + \epsilon^\mu$, $\delta\xi^a = \epsilon^\mu \partial_\mu \xi^a$, by using \eqref{G1}, we find 
\begin{align}
\label{G1GCT}
\delta \Gamma^\rho_{\mu\nu}
= \epsilon^\sigma \partial_\sigma \Gamma^\rho_{\mu\nu} - \partial_\sigma \epsilon^\rho \Gamma^\sigma_{\mu\nu} 
+ \partial_\mu \epsilon^\eta \Gamma^\rho_{\eta\nu} + \partial_\nu \epsilon^\eta \Gamma^\rho_{\mu\eta} 
+ \partial_\mu \partial_\nu \epsilon^\rho \, . 
\end{align}
The last term in \eqref{G1GCT} is nothing but the inhomogeneous term, which is specific to the connection and keeps the general covariance of the covariant derivative. 
This may also tell us that $\xi^a$'s are surely scalar fields. 

We define a non-metricity scalar
\begin{align}
\label{Q}
Q=&\, - \frac{1}{4} g^{\alpha\mu} g^{\beta\nu} g^{\gamma\rho} \nabla_\alpha g_{\beta\gamma} \nabla_\mu g_{\nu\rho}
+ \frac{1}{2} g^{\alpha\mu} g^{\beta\nu} g^{\gamma\rho} \nabla_\alpha g_{\beta\gamma} \nabla_\rho g_{\nu\mu} \nonumber \\
&\, + \frac{1}{4} g^{\alpha\mu} g^{\beta\gamma} g^{\nu\rho} \nabla_\alpha g_{\beta\gamma} \nabla_\mu g_{\nu\rho} 
 - \frac{1}{2} g^{\alpha\mu} g^{\beta\gamma} g^{\nu\rho} \nabla_\alpha g_{\beta\gamma} \nabla_\nu g_{\mu\rho} \, .
\end{align}
The difference between $Q$ and the scalar curvature $\tilde R$ of Einstein's gravity is a total derivative, 
\begin{align}
\label{RQ}
\tilde R = Q - {\tilde\nabla}_\alpha\left(Q^{\alpha}-\tilde{Q}^{\alpha}\right)\, .
\end{align} 
Here $Q_\sigma \equiv Q^{\ \mu}_{\sigma\ \mu}$, $\tilde{Q}_\sigma=Q^\mu_{\ \sigma \mu}$. 
This tells us that the action linear to $Q$ is equivalent to the Einstein-Hilbert action in Einstein's gravity. 

The action of $f(Q)$ gravity with a function $f(Q)$ of $Q$ is given by,
\begin{align}
\label{fQactn} 
S=\int d^4 x \sqrt{-g} f(Q)\, ,
\end{align}
which is different from the action of $f(R)$ gravity. 

We regard the metric $g_{\mu\nu}$ and $\xi^a$ as independent fields hereafter, as proposed in \cite{Nojiri:2024zab}.

For the $f(Q)$ gravity, there have been presented arguments about the dynamical degrees of freedom (DOF). 
In \cite{Heisenberg:2023wgk}, it has been shown that there are up to seven degrees of freedom. 
After that, in \cite{Gomes:2023tur}, it has been claimed that there could exist seven degrees of freedom. 
The existence of the ghost was also claimed because the conformal mode of the metric is a ghost. 
Before the paper \cite{Gomes:2023tur}, however, we have shown that the conformal mode was a ghost, but it does not propagate due to the constraint~\cite{Hu:2023gui}. 
Therefore, the seven degrees of freedom could not be ghosts. 
Hence, the number of the dynamical degrees of freedom has not been clarified.

\section{Cosmological Reconstruction of $f(Q)$ Gravity\label{Sec3}}

We now consider how one can reconstruct the realistic universe expansion in the spatially flat FLRW spacetime 
\begin{align}
\label{FLRW}
ds^2=-dt^2+a^2(t)\sum_{i=1,2,3} \left(dx^i\right)^2 \, .
\end{align}
Here, $a(t)$ is called a scale factor. 
In the FLRW spacetime~\eqref{FLRW}, the metricity scalar in \eqref{Q} has the following form, 
\begin{align}
\label{QFLRW}
Q = - 6 H^2 \, .
\end{align}
Here, the Hubble rate $H$ is defined by $H=\frac{\dot a}{a}$. 

We write the equation corresponding to the Einstein equation in the following form, 
\begin{align}
\label{EQeq}
0= \mathcal{G}_{\mu\nu} + T_{\mu\nu}\, .
\end{align}
Then the equations corresponding to the Friedmann equations in Einstein's gravity are given by 
\begin{align}
\label{vQ_FLRW0}
- \rho =&\, \mathcal{G}_{00} = - \frac{f(Q)}{2} - 6 H^2 f'(Q) \, , \nonumber \\
 - p a^2 \delta_{ij} =&\, \mathcal{G}_{ij} = a^2 \delta_{ij} \left\{ \frac{f(Q)}{2} + 2 a^{-3} \frac{d}{dt} \left( a^3 H f'(Q) \right)\right\} \nonumber \\
=&\, a^2 \delta_{ij} \left\{ \frac{f(Q)}{2} + 6H^2 f'(Q) + 2 \dot H f'(Q) - 24 H^2 \dot H f''(Q) \right\} \, .
\end{align}
Here $\rho$ and $p$ are the energy density and the pressure of matter, respectively. 

One may check the conservation law of matter, 
\begin{align}
\label{cons}
\dot\rho + 3 H \left( \rho + p \right) = 0 \, .
\end{align}
The conservation of matter can also be obtained only from the matter equation of motion. 
The conservation law of the matter energy-momentum tensor $T_{\mu\nu}$ holds regardless of what kind of gravity theory is considered. 
Therefore the conservation should be described by the Levi-Civita connection $\tilde \Gamma^\sigma_{\;\mu \nu}$ in \eqref{Levi-Civita} of Einstein's gravity even if we consider the $f(Q)$ gravity theory, 
\begin{align}
\label{cons2}
0 = \tilde\nabla^\mu T_{\mu\nu} = g^{\rho\mu} \left( \partial_\rho T_{\mu\nu} - \tilde \Gamma^\sigma_{\; \rho\mu} T_{\sigma\nu} 
 - \tilde \Gamma^\sigma_{\; \rho\nu} T_{\sigma\mu} \right) \, .
\end{align}
As we are considering the theory with action, the consistency of the Lagrangian theory, that is, a condition of functional integrability, shows 
\begin{align}
\label{QBianchi}
0 = \tilde\nabla^\mu \mathcal{G}_{\mu\nu} \, .
\end{align}
This gives a generalised Bianchi identity in $f(Q)$ gravity theory. 

In the case of $f(Q)$ gravity, there is an exceptional model, where arbitrary development of the universe expansion in the FLRW spacetime~\eqref{FLRW} is a solution when we neglect the contribution of the matter. 
Then we obtain, 
\begin{align}
\label{vQ_FLRW0vacuum}
0 =&\, \mathcal{G}_{00} = - \frac{f(Q)}{2} - 6 H^2 f'(Q) \, , \nonumber \\
0 =&\, \mathcal{G}_{ij} = a^2 \delta_{ij} \left\{ \frac{f(Q)}{2} + 6H^2 f'(Q) + 2 \dot H f'(Q) + 24 H^2 \dot H f''(Q) \right\} \, .
\end{align}
Because Eq.~\eqref{cons} also shows that in the vacuum, where $\rho=p=0$, the second equation in \eqref{vQ_FLRW0vacuum} can be obtained from the first equation, we forget the second equation. 
By using Eq.~\eqref{QFLRW}, we rewrite the first equation in \eqref{vQ_FLRW0vacuum} in the following form, 
\begin{align}
\label{vQ_FLRW1}
0 = - \frac{1}{2} f(Q) + Q f'(Q) \, .
\end{align}
Then we find that $f(Q)$ is given by 
\begin{align}
\label{vQ_FLRW2}
f(Q) = f_0 \sqrt{ - Q } \, .
\end{align}
Therefore, if we consider the model \eqref{vQ_FLRW2}, the arbitrary development of the universe expansion is a solution. 

We now consider the case that there are several kinds of matter. 
We assume the matter satisfies the conservation law \eqref{cons}. 
The conservation law determines the scale dependence of $\rho$ and $p$, $\rho=\rho(a)$ and $p=p(a)$. 
If the scale factor $a=a(t)$ is given, we find the $t$ dependences of $\rho$, $p$, and $Q$, $\rho=\rho(t)$, $p=p(t)$, and $Q=-6H^2 = Q(t)$. 
By solving $Q=Q(t)$ with respect to $t$, $t=t(Q)$, and by substituting the obtained expression into $\rho=\rho(t)$ and $p=p(t)$, 
we find $Q$ dependence of $\rho$ and $p$ as $\rho=\rho(Q)$ and $p=p(Q)$. 
We rewrite the first equation in \eqref{vQ_FLRW0} as follows, 
\begin{align}
\label{vQ_FLRW0Q}
- \rho (Q) = - \frac{f(Q)}{2} + Q f'(Q) = - \left(-Q\right)^\frac{3}{2} \frac{d}{dQ} \left( \frac{f(Q)}{\left(-Q\right)^\frac{1}{2}} \right) \, .
\end{align}
The solution of \eqref{vQ_FLRW0Q} is given by 
\begin{align}
\label{arbQ}
f(Q)= \left( -Q \right)^\frac{1}{2} \int^Q dq \rho(q) \left( - q \right)^{-\frac{3}{2}} \, .
\end{align}
By using \eqref{arbQ}, one can find the $f(Q)$ gravity corresponding to the arbitrary expansion history of the universe given by $a=a(t)$. 

Matter could be created at the end of inflation by the quantum corrections. 
These corrections are not included in the classical action. 
As an effective theory, we may include the effects by modifying \eqref{vQ_FLRW0} as follows, 
\begin{align}
\label{vQ_FLRW0_eff}
\frac{f(Q)}{2} + 6 H^2 f'(Q) =&\, \rho_\mathrm{eff} \equiv \rho + \mathcal{J}_\rho (Q) \, , \nonumber \\
 - \left\{ \frac{f(Q)}{2} + 2 a^{-3} \frac{d}{dt} \left( a^3 H f'(Q) \right)\right\} =&\, p_\mathrm{eff} \equiv p + \mathcal{J}_p(Q)\, \, .
\end{align}
Here $\mathcal{J}_\rho(Q)$ and $\mathcal{J}_p(Q)$ are some functions of $Q$. 
Eq.~\eqref{cons2} tells that $\rho_\mathrm{eff}$ and $p_\mathrm{eff}$ satisfy the conservation law as in \eqref{cons}, which gives, 
\begin{align}
\label{conseff}
\dot\rho + 3 H \left( \rho + p \right) = J \equiv - \dot Q \mathcal{J}_\rho'(Q) - 3 H \left( \mathcal{J}_\rho(Q) + \mathcal{J}_p(Q) \right) \, .
\end{align}
Therefore, we find that $J$ can act as a source of matter. 
We may choose $J$ so that $J$ does not vanish only at the end of the inflation and generates matter. 
In general, the functions $\mathcal{J}_\rho$ and $\mathcal{J}_p$ can depend on $\dot Q$, $\ddot Q$, $\cdots$ in addition to $Q$ but just for simplicity, we choose $\mathcal{J}_\rho$ and $\mathcal{J}_p$ to be functions of only $Q$. 

\section{Models of inflation\label{Sec4}}

We now consider the inflation by using Eq.~\eqref{arbQ}. 

Because radiation dominates after inflation, we consider only radiation as matter. 
In the radiation-dominated universe, the Hubble rate $H$ behaves as $H\sim \frac{1}{2t}$ and during the inflation, of course, $H=H_0$ $\left(H_0:\ \mbox{constant}\right)$. 
A unifying behaviour is given by 
\begin{align}
\label{Hex1b}
H(t) = \frac{H_0}{1 + \alpha \ln \left( 1 + \e^{\frac{2 H_0}{\alpha} \left( t - t_0 \right)} \right)} \, .
\end{align}
Here, $\alpha$ is a positive constant and $t_0$ is a constant which corresponds to the time when the inflation ends. 
When $t\ll t_0$, we obtain that $H$ becomes a constant $H\to H_0$, which corresponds to the inflation. 
On the other hand, when $t\gg t_0$, we find $H\to \frac{1}{2\left(t - t_0 \right)}$, which corresponds to the radiation-dominated universe. 

The slow-roll parameters $\epsilon$ and $\eta$ are defined as 
\begin{align}
\label{slwrllprmtrs}
\epsilon = - \frac{\dot H}{H^2}\, , \quad \eta = \frac{\dot{\epsilon}}{\epsilon H}\, .
\end{align}
The end of the inflation may be defined by $\epsilon=1$. 
For the model \eqref{Hex1b}, we find 
\begin{align}
\label{slowroll2}
\epsilon = \frac{2}{1 + \e^{- \frac{2 H_0}{\alpha} \left( t - t_0 \right)}} \, , \quad 
\eta = \frac{2\left\{ 1 + \alpha \ln \left( 1 + \e^{\frac{2 H_0}{\alpha} \left( t - t_0 \right)} \right) \right\}
\e^{- \frac{2 H_0}{\alpha} \left( t - t_0 \right)}}{\alpha \left( 1 + \e^{- \frac{2 H_0}{\alpha} \left( t - t_0 \right)}\right)}\, .
\end{align}
Therefore, when $t=t_0$, we surely find $\epsilon=1$. 

Because $Q=-{ 6}H^2$, one obtains 
\begin{align}
\label{Hex1b2}
\e^{\frac{2 H_0}{\alpha} \left( t - t_0 \right)} = \e^{\frac{1}{\alpha} \left(\sqrt{- \frac{{ 6}{H_0}^2}{Q}} - 1\right)} - 1 \quad \mbox{or} \quad 
t=t_0 + \frac{\alpha}{2H_0} \ln \left( \e^{\frac{1}{\alpha} \left( \sqrt{- \frac{{ 6}{H_0}^2}{Q}} - 1\right) } - 1 \right) \, .
\end{align}
Therefore, $t$ is explicitly given as a function of $Q$. 

We now find 
\begin{align}
\label{derivatives}
\dot H = - \frac{2 H^2}{1 + \e^{-\frac{2 H_0}{\alpha} \left( t - t_0 \right)}} 
= 2 H^2 \left( 1 - \e^{-\frac{1}{\alpha} \left(\frac{H_0}{H} - 1 \right)} \right)
= { - \frac{Q}{3}} \left( 1 - \e^{-\frac{1}{\alpha} \left(\sqrt{- \frac{{ 6}{H_0}^2}{Q}}-1\right)} \right)\, , 
\end{align}
which will be used later. 

As it was mentioned around Eq.~\eqref{conseff}, we may assume the radiation is generated at the end of the inflation $t\sim t_0$. 
One may assume the behaviour of the energy density $\rho$ of the radiation as follows, 
\begin{align}
\label{Hex1b3}
\rho = - \frac{Q \left( Q + 6 {H_0}^2 \right)}{2\kappa^2 \left( - Q + 6 {H_0}^2 \right)} 
= - \frac{1}{2\kappa^2} \left( - Q - 12 {H_0}^2 + \frac{72{H_0}^4}{ - Q + 6{H_0}^2} \right) \, .
\end{align}
When $t\ll t_0$, because $Q= -{ 6}H^2 \to - { 6}{H_0}^2$, we obtain $\rho \to 0$. 
On the other hand, when $t\gg t_0$, because $Q\ll { 6}{H_0}^2$, $\rho$ behaves as $\rho\to - \frac{3Q}{{ 6}\kappa^2} = \frac{3H^2}{\kappa^2}$ as expected in Einstein's gravity. 
Because the equation of state (EoS) parameter of the radiation is $\frac{1}{3}$, the pressure $p$ is given by 
\begin{align}
\label{Hex1b3p}
p = - \frac{Q \left( Q + 6{H_0}^2 \right)}{6 \kappa^2 \left( - Q + 6{H_0}^2 \right)} \, .
\end{align}
Note that the above $\rho$ and $p$ do not satisfy the conservation law \eqref{cons} but violate it as in \eqref{conseff}. 
In fact, by using \eqref{derivatives}, we find, 
\begin{align}
\label{consmod1}
J=&\, \dot \rho + 3 H \left( \rho + p \right) 
= \frac{H}{\kappa^2} \left\{ \frac{ 2 Q \left( - Q^2 + 12 {H_0}^2 Q + 36 {H_0}^4 \right)}{ \left( - Q + 6{H_0}^2\right)^2 } 
\left( 1 - \e^{-\frac{1}{\alpha} \left( \sqrt{- \frac{6{H_0}^2}{Q}}-1\right)} \right) 
 - \frac{Q \left( Q + 6{H_0}^2 \right)}{ - Q + 6{H_0}^2 } \right\} \, ,
\end{align}
which vanishes at the early time $t\ll t_0$ or $Q\to - { 6}{H_0}^2$ and at the late time $t\gg t_0$ or $Q\to 0$. 
The source $J$ vanishes except at the end of the inflation $t\sim t_0$, when the radiation is generated. 
There are some ambiguities in choosing $\mathcal{J}_\rho(Q)$ in \eqref{vQ_FLRW0_eff}. 
For simplicity, one may choose $\mathcal{J}_\rho(Q)=0$. 
Then Eq.~\eqref{conseff} gives, 
\begin{align}
\label{fp}
\mathcal{J}_p = - \frac{1}{3\kappa^2} \left\{ \frac{ 2 Q \left( - Q^2 + 12 {H_0}^2 Q + 36 {H_0}^4 \right)}{ \left( - Q + 6{H_0}^2\right)^2 } 
\left( 1 - \e^{-\frac{1}{\alpha} \left( \sqrt{- \frac{6{H_0}^2}{Q}}-1\right)} \right) 
 - \frac{Q \left( Q + 6{H_0}^2 \right)}{ - Q + 6{H_0}^2 } \right\} \, ,
\end{align}
which may effectively express the quantum generation of the radiation. 

By using \eqref{arbQ}, we find the explicit form of $f(Q)$, 
\begin{align}
\label{arbQinf}
f(Q)
=&\, \frac{1}{2\kappa^2} \left\{ - 2 Q + 24 {H_0}^2 + 4 \left(6 {H_0}^2\right)^\frac{1}{2} \left( -Q \right)^\frac{1}{2} 
\mathrm{Arctan} \left( \left( \frac{-Q}{6 {H_0}^2} \right)^\frac{1}{2} \right)
 - C \left( -Q \right)^\frac{1}{2} \right\} \, .
\end{align}
Here, $C$ is a constant of integration
Note that we have chosen $\rho_\mathrm{eff}=\rho$. 

We may also consider another model without introducing $\mathcal{J}_\rho$ or $\mathcal{J}_p$. 
The model, therefore, describes the universe until the end of inflation. 
For the model, we use the $e$-folding $N$, defined in terms of the scale factor $a$ as $a=\e^{N-N_0}$. 
The constant $N_0$ is introduced so that $a=1$ in the present universe, although we now consider the universe until the end of inflation. 

In the model, the Hubble rate $H$ is assumed to be, 
\begin{align}
\label{HubbleNewC}
H = \frac{H_0}{\left(1 + \e^{\beta\left(N-N_1\right)}\right)^\gamma} \, .
\end{align}
Here $H_0$, $N_1$, $\beta$ and $\gamma$ are constants. 
When $N-N_1$ is large and negative, we find $H\sim H_0$, which corresponds to the inflation, 
On the other hand, when $N-N_1$ is large and positive, $H$ decreases as $H \sim H_0 \e^{-\beta\gamma\left(N-N_1\right)}$, which may correspond to the end of the inflation. 

For the model \eqref{HubbleNewC}, the slow-roll parameters are given by, 
\begin{align}
\label{slwrllprmtrsNewC}
\epsilon = \frac{\beta\gamma \e^{\beta\left(N-N_1\right)}}{1 + \e^{\beta\left(N-N_1\right)}} \, , \quad 
\eta = \frac{\beta}{1 + \e^{\beta\left(N-N_1\right)}} \, .
\end{align}
This tells that the end of the inflation $\epsilon=1$ is given by the following value of $N$
\begin{align}
\label{endN}
N=N_1 = \frac{1}{\beta} \ln \frac{1}{\beta\gamma-1} \, .
\end{align}
So that $N$ exists, we find the following constraint on the parameters, 
\begin{align}
\label{cnstrntbtgmm}
\beta\gamma >1 \, .
\end{align}

Eq.~\eqref{HubbleNewC} gives, 
\begin{align}
\label{Na}
N= N_1 + \frac{1}{\beta} \ln \left( \left( \frac{H_0}{H} \right)^\frac{1}{\gamma} - 1 \right) \, , \quad 
a= \e^{N_1-N_0} \left( \left( \frac{H_0}{H} \right)^\frac{1}{\gamma} - 1 \right)^\frac{1}{\beta}
= \e^{N_1-N_0} \left( \left( H_0 \sqrt{ - \frac{6}{Q}} \right)^\frac{1}{\gamma} - 1 \right)^\frac{1}{\beta} \, .
\end{align}
If the matter during the inflation is only a small radiation, we find 
\begin{align}
\label{rhoC}
\rho = \rho_0 a^{-4} = \rho_0 \e^{4 \left(N_1-N_0\right) } \left( \left( H_0 \sqrt{ - \frac{6}{Q}} \right)^\frac{1}{\gamma} - 1 \right)^\frac{4}{\beta} \, ,
\end{align}
which gives the form of $f(Q)$ by using \eqref{arbQ}, 
\begin{align}
\label{arbCQ}
f(Q) =&\, 2 \gamma \rho_0 \e^{4 \left(N_1-N_0\right)} \left( -\frac{Q}{6{H_0}^2} \right)^\frac{1}{2} \left\{
B_{\left(-\frac{Q}{6{H_0}^2}\right)^\frac{1}{2\gamma}} \left( 3\gamma - \frac{4}{\beta}, \frac{4}{\beta} + 1 \right) 
+ C \right\}\, .
\end{align}
Here $C$ is a constant of the integration, again, and $B_z(a,b)$ is the incomplete beta function defined by 
\begin{align}
\label{beta}
B_z(a,b)=\int^z_0 dt t^{a-1} \left( 1 - t\right)^{b-1} \, .
\end{align}
Then we find the explicit form of $f(Q)$, which may describe inflation. 
As one can see, the number of such theories is restricted by the above forms. 

\section{$f(Q)$ gravity satisfying the ACT-DR6 constraints\label{Sec5}}

The ACT-DR6 gives some constraints on the parameters of inflation, like the spectral index parameter $n_s$ and the tensor-to-scalar ratio parameter $r$ \cite{AtacamaCosmologyTelescope:2025blo, AtacamaCosmologyTelescope:2025nti}. 
When we assume that we observe the cosmic microwave radiation (CMB) with the wavenumber $k\sim 0.05\, \mathrm{Mpc}^{-1}$, the emitted time $t_*$ is estimated to be 
\begin{align}
\label{k}
k=a(t_*) H(t_*)\, .
\end{align}
{
In the case of the scalar-tensor theory and $F(R)$ gravity theory~\cite{Hwang:1990re, Hwang:1990jh, Makino:1991sg}, by} using the estimated $t_*$ and the slow-roll parameters $\epsilon$ and $\eta$ defined in \eqref{slwrllprmtrs} when $t=t_*$, the spectral index parameter $n_s$ and the tensor-to-scalar ratio parameter $r$ are given by 
\begin{align}
\label{prmtrs}
n_s - 1 = - 2 \epsilon(t_*) - \eta(t_*)\, , \quad r = 12 \epsilon (t_*)\, .
\end{align}
{
The above expressions have been well-established~\cite{Hwang:1990re, Hwang:1990jh, Makino:1991sg} for the scalar-tensor theory and $F(R)$ gravity theory and they are used very well. 
We now use the expressions~\eqref{prmtrs} even for $f(Q)$ gravity, which is an assumption, or a working hypothesis, that needs more detailed discussion, although the matter fluctuations could be determined solely by the background evolution given by $H$. 
As shown in \cite{Capozziello:2024vix}, the only propagating mode in the flat background is a graviton at least perturbatively. 
If this holds even in the FLRW universe~\eqref{FLRW}, there is no scalar mode corresponding to the inflaton, so we need to find the origin of the scalar perturbation. 
In the flat background, the perturbation $\delta Q$ of $Q$ gives a non-trivial contribution in the quadratic order of the perturbation of the metric and connections because the non-metricity scalar$Q$ is given by the square of the non-metricity tensor defined in \eqref{non-metricity} as in \eqref{Q}, or the square of the covariant derivative of the metric. 
In the FLRW universe, because the covariant derivative of the metric does not vanish, $\delta Q$ includes the linear term of the perturbation of the metric, which generates terms like $\frac{1}{2} f''\left( \bar Q \right) \left( \delta Q \right)^2$ in the perturbed Lagrangian density. 
Here $\bar Q$ is $Q$ in the background. 
This term may generate the scalar perturbation as in \cite{BeltranJimenez:2019tme}. 
The matter included, as in \eqref{arbQ}, may also generate the scalar perturbation. 
We should also note that even for the tensor perturbation, there could be some corrections coming from $f''\left( \bar Q \right)$. 
The last term $C \left( -Q \right)^\frac{1}{2}$ in \eqref{arbQinf}, which always appear acompany with the constant of integration $C$, contributes to $f''\left( \bar Q \right)$. 
Then by tuning $C$, we may adjust $f''\left( \bar Q \right)$, for example, we may choose $C$ so that $f''\left( \bar Q \right)\sim 0$ when $t\sim t_*$. 
Anyway, we need much more detailed arguments to determine more exact expressions for the spectral index parameter $n_s$ and the tensor-to-scalar ratio parameter $r$, but in this paper, we continue the arguments by following the simple assumption \eqref{prmtrs}. 
If there is no consistent solution even if we use \eqref{prmtrs}, we may need to discard $f(Q)$ theory as a model of inflation. 
A possibility is that $f(Q)$ gravity might be a piece of the gravity theory like $R^2$-model, where the contribution of the $f(Q)$ gravity to the spectral index parameter $n_s$ and the tensor-to-scalar ratio parameter $r$ is not main but the $f(Q)$ part make a supplementary contribution. 
}

For the model \eqref{Hex1b}, by using \eqref{slowroll2}, we find, 
\begin{align}
\label{index00}
n_s = 1 - \frac{4\left[ 1 + \frac{1}{\alpha} \left\{ 1 + \alpha \ln \left( 1 + \e^{\frac{2 H_0}{\alpha} \left( t_* - t_0 \right)} \right) \right\}
\e^{-\frac{2 H_0}{\alpha} \left( t_* - t_0 \right)}\right]}
{1 + \e^{- \frac{2 H_0}{\alpha} \left( t_* - t_0 \right)}} \, , \quad 
r = \frac{32}{1 + \e^{-\frac{2 H_0}{\alpha} \left( t_* - t_0 \right)}} \, .
\end{align}
We now check whether the model \eqref{Hex1b} is consistent with the ACT constraint. 

The ACT observation~\cite{AtacamaCosmologyTelescope:2025nti} gives the constraint on the value of $n_s$, 
\begin{align}
\label{nsACTonly}
n_s = 0.974 \pm 0.009\, .
\end{align}
Only by the ACT observation, however, there is no constraint on the value of $r$ because the ACT does not directly observe the low $l$ B-mode polarisation of the CMB ($l\leq 100$). 
The constraint on $r$ is mainly obtained from the BICEP/Keck observation. 
The combined data of Planck, ACT, lensing, BAO, BK18 (BICEP/Keck) tell \cite{AtacamaCosmologyTelescope:2025nti} the medians of $n_s$ and $r$ as follows, 
\begin{align}
\label{nsr}
n_s \sim 0.976 - 0.977\, , \quad r \sim 0.012 \ \mbox{or}\ r<0.038 \, .
\end{align}
{
We should note, however, that recently, it is pointed out that the values of $n_s$ are not stable against the assumed pre-recombination model, as well as the primordial power spectrum parametrisation \cite{Sabogal:2026qvy}. 
}

If $r=0.012$, by using the expression of $r$ in \eqref{index00}, we find 
\begin{align}
\label{r} 
\e^{-\frac{2 H_0}{\alpha} \left( t_* - t_0 \right)} \sim 2.7\times 10^3 \, .
\end{align}
Then the expression of $n_s$ gives 
\begin{align}
\label{ns}
n_s \sim 1 - \frac{4\left[ 1 + \frac{1}{\alpha} \left\{ 1 + \frac{\alpha}{2.7\times 10^3} \right\} 2.7\times 10^3 \right]}{2.7\times 10^3} \, .
\end{align}
If we choose $n_s=0.976$, we find 
\begin{align}
\label{ns2}
0.024 \sim \frac{4\left[ 1 + \frac{1}{\alpha} \left\{ 1 + \frac{\alpha}{2.7\times 10^3} \right\} 2.7\times 10^3 \right]}{2.7\times 10^3} \, , 
\end{align}
that is, 
\begin{align}
\label{alph}
\alpha \sim 1.9\times 10^2 \, .
\end{align}
The condition \eqref{k} could be satisfied by adjusting $H_0$ and $t_0$. 
Hence, we obtain a model that satisfies \eqref{nsr} and is consistent with the data via ACT, etc. 
We should note $\alpha$ appears in \eqref{fp}. 

Then a model satisfying the constraints coming from the ACT observation, etc., is given by \eqref{arbQinf} and \eqref{fp}
\begin{align}
\label{arbQinfFF}
f(Q) 
=&\, \frac{1}{2\kappa^2} \left\{ - 2 Q + 24 {H_0}^2 + 4 \left(6 {H_0}^2\right)^\frac{1}{2} \left( -Q \right)^\frac{1}{2} 
\mathrm{Arctan} \left( \left( \frac{-Q}{6 {H_0}^2} \right)^\frac{1}{2} \right)
 - C \left( -Q \right)^\frac{1}{2} \right\} 
\, , \nonumber \\
\mathcal{J}_\rho =&\, 0 \, , \nonumber \\ 
\mathcal{J}_p =&\, - \frac{1}{3\kappa^2} \left\{ \frac{ 2 Q \left( - Q^2 + 12 {H_0}^2 Q + 36 {H_0}^4 \right)}{ \left( - Q + 6{H_0}^2\right)^2 } 
\left( 1 - \e^{-\frac{1}{\alpha} \left( \sqrt{- \frac{6{H_0}^2}{Q}}-1\right)} \right) 
 - \frac{Q \left( Q + 6{H_0}^2 \right)}{ - Q + 6{H_0}^2 } \right\} \, ,
\end{align}
with \eqref{r} and \eqref{alph}, 
\begin{align}
\label{pp}
\e^{-\frac{2 H_0}{\alpha} \left( t_* - t_0 \right)} \sim 2.7\times 10^3 \, , \quad 
\alpha \sim 1.9\times 10^2 \, .
\end{align}
Then the above $f(Q)$ gravity inflation is consistent with the observations. 

For the model \eqref{HubbleNewC}, because we are using the $e$-folding number $N$, Eq.~\eqref{k} and the definitions of the spectral index parameter $n_s$ and the tensor-to-scalar ratio parameter $r$ in \eqref{prmtrs} are expressed by 
\begin{align}
\label{kN}
k=&\, a(N_*) H(N_*)\, , \\
\label{prmtrsN}
n_s - 1 =&\, - 2 \epsilon(N_*) - \eta(N_*)\, , \quad r = 12 \epsilon (N_*)\, .
\end{align}
Here $N_*$ is the $e$-folding number corresponding to the observed emission of CMB. 

By using the expressions of the slow-roll parameters $\epsilon$ and $\eta$ in \eqref{slwrllprmtrsNewC}, we find the following expressions of the spectral index parameter $n_s$ and the tensor-to-scalar ratio parameter $r$, 
\begin{align}
\label{prmtrsNewC}
n_s = 1 - \frac{\beta\left( 1 + 2\gamma \e^{\beta\left(N_*-N_1\right)}\right)}{1 + \e^{\beta\left(N_*-N_1\right)}} \, , \quad 
r = \frac{12\beta\gamma \e^{\beta\left(N_*-N_1\right)}}{1 + \e^{\beta\left(N_*-N_1\right)}} \, .
\end{align}
By the combined data \eqref{nsr}, we choose $n_s=0.976$ and $r=0.012$, again. 
Then the equations \eqref{prmtrsNewC} tell, 
\begin{align}
\label{nsr2}
0.024 = \frac{\beta\left( 1 + 2\gamma \e^{\beta\left(N_*-N_1\right)}\right)}{1 + \e^{\beta\left(N_*-N_1\right)}} \, , \quad 
0.012 = \frac{12\beta\gamma \e^{\beta\left(N_*-N_1\right)}}{1 + \e^{\beta\left(N_*-N_1\right)}} \, ,
\end{align}
that is, 
\begin{align}
\label{betagamma}
\beta \sim 0.023 \times \left( 1 + \e^{\beta\left(N_*-N_1\right)} \right) \, , \quad 
\gamma \sim 0.43 \times \e^{-\beta\left(N_*-N_1\right)} \, .
\end{align}
On the other hand, the condition \eqref{kN} shows
\begin{align}
\label{kNC}
k = \frac{H_0 \e^{N_1-N_0}}{\left(1 + \e^{\beta\left(N_*-N_1\right)}\right)^\gamma} \, .
\end{align}
Because we expect $H_0\sim 10^{14}\, \mathrm{GeV}=10^{23}\, \mathrm{eV}$ and $k\sim 0.05\, \mathrm{Mpc}^{-1}$, we find $\e^{N_1-N_0}\sim 10^{-54}$. 
Eq.~\eqref{kNC} shows that $N_*$ is given by 
\begin{align}
\label{Nstar}
N_*= N_1 + \frac{1}{\beta} \ln \left( \left(\frac{H_0 \e^{N_1-N_0}}{k}\right)^\frac{1}{\gamma} - 1 \right) \, .
\end{align}
We also find 
\begin{align}
\label{betaNstar}
\e^{\beta\left(N_*-N_1\right)} = \left(\frac{H_0 \e^{N_1-N_0}}{k}\right)^\frac{1}{\gamma} - 1 \, .
\end{align}
This may tell that $\e^{\beta\left(N_*-N_1\right)} = \mathcal{O} (1)$. 
Anyway, there are solutions for the parameters $\beta$, $\gamma$, which satisfy the constraint \eqref{nsr} coming from the observations of ACT. 
An explicit form of the model is given by \eqref{arbCQ}, 
\begin{align}
\label{arbCQB}
f(Q)
=&\, 2 \gamma \rho_0 \e^{4 \left(N_1-N_0\right)} \left( -\frac{Q}{6{H_0}^2} \right)^\frac{1}{2} \left\{
B_{\left(-\frac{Q}{6{H_0}^2}\right)^\frac{1}{2\gamma}} \left( 3\gamma - \frac{4}{\beta}, \frac{4}{\beta} + 1 \right) + C \right\} \, , \nonumber \\
&\, \beta = 0.023 \, , \quad \gamma = 86 \, , 
\end{align}
which gives $\e^{\beta\left(N_*-N_1\right)} \sim 0.5\times 10^{-2}$ and satisfies the constraint \eqref{cnstrntbtgmm}. 

\section{$f(Q)$ gravity coupled with a scalar field and cosmology\label{SecVI}}

In this section, as an extension of the $f(Q)$ gravity, we consider the model where the scalar field $\phi$ is minimally coupled with $f(Q)$ gravity. 
The action, which we consider, is the following, 
\begin{align}
\label{fQactn}
S=\int d^4 x \sqrt{-g} \left( f(Q) - \frac{1}{2}\omega(\phi) g^{\mu\nu} \partial_\mu \phi \partial_\nu \phi - V(\phi) \right) \, .
\end{align}
Here $V(\phi)$ is the potential of the scalar function, and we also introduce another function $\omega(\phi)$ for later convenience. 
In this section, we neglect the contribution from matter, although it is straightforward to include the matter. 
We consider the formulation as in \cite{Nojiri:2005pu, Capozziello:2005tf}

We assume the spatially flat FLRW spacetime in \eqref{FLRW}, again, and we also assume that the scalar field $\phi$ only depends on the cosmological time $t$. 
Then the Friedmann equations corresponding to \eqref{vQ_FLRW0} have the following form 
\begin{align}
\label{vQ_FLRW0phi}
\frac{1}{2} \omega(\phi) {\dot\phi}^2 + V(\phi) =&\, \frac{f(Q)}{2} - Q H^2 f'(Q) \, , \nonumber \\
\frac{1}{2} \omega(\phi) {\dot\phi}^2 - V(\phi) =&\, - \frac{f(Q)}{2} + Q f'(Q) - 2 \dot H f'(Q) - 4 Q \dot H f''(Q) \, .
\end{align}
Because the redefinition of the scalar field can be absorbed into the function $\omega(\phi)$, we may identify the scalar field with the cosmological time, $\phi=t$. 
By the identification, the equations in \eqref{vQ_FLRW0phi} are simplified to be 
\begin{align}
\label{vQ_FLRW0phismpl}
\frac{1}{2} \omega(t) + V(t) =&\, \frac{f(Q)}{2} - Q f'(Q) \, , \nonumber \\
\frac{1}{2} \omega(t) - V(t) =&\, - \frac{f(Q)}{2} + Q f'(Q) - 2 \dot H f'(Q) - 4 Q \dot H f''(Q) \, ,
\end{align}
which are solved with respect to $\omega$ and $V$ as follows, 
\begin{align}
\label{vQ_FLRW0phiomgV}
\omega(t) =&\, - 2 \dot H f'(Q) - 4 Q \dot H f''(Q) \, , \nonumber \\
V(t) =&\, \frac{f(Q)}{2} - Q f'(Q) - \dot H f'(Q) - 2 Q \dot H f''(Q) \, .
\end{align}
This tells that by using a function $\eta(\phi)$, if we consider the model, 
\begin{align}
\label{vQ_FLRW0eta}
\omega(\phi) =&\, - 2\eta'(\phi) f'\left( - 6\eta(\phi)^2 \right) + 24 \eta(\phi)^2 \eta'(\phi) f''\left( - 6 \eta(\phi)^2 \right) \, , \nonumber \\
V(\phi) =&\, \frac{f\left( - 6 \eta(\phi)^2 \right)}{2} + 6 \eta(\phi)^2 f'\left( - 6 \eta(\phi)^2 \right) - \eta'(\phi) f'\left( -6 \eta(\phi)^2 \right) 
+ 12 \eta(\phi)^2 \eta'(\phi) f''\left( - 6 \eta(\phi)^2 \right) \, ,
\end{align}
an exact solution is given by 
\begin{align}
\label{sol}
\phi=t \, , \quad H=\eta(t)\, .
\end{align}
Here $f(Q)$ is almost arbitrary, but in \eqref{vQ_FLRW0eta}, we need to choose $f(Q)$ so that a ghost does not appear. 
Then, if we choose 
\begin{align}
\label{Hex1beta}
\eta(\phi) = \frac{H_0}{1 + \alpha \ln \left( 1 + \e^{\frac{2 H_0}{\alpha} \left( \phi - t_0 \right)} \right)} \, .
\end{align}
The Hubble rate $H$ in \eqref{Hex1b} is reproduced, and we obtain a model consistent with the ACT observations. 

Instead of identifying $\phi$ with the cosmological time $t$, we may identify $\phi$ with the $e$-folding number $N$. 
Then instead of \eqref{vQ_FLRW0phismpl}, we find, 
\begin{align}
\label{vQ_FLRW0phismplN}
\frac{1}{2} H^2 \omega(N) + V(N) =&\, \frac{f(Q)}{2} - Q f'(Q) \, , \nonumber \\
\frac{1}{2} H^2 \omega(N) - V(N) =&\, - \frac{f(Q)}{2} + Q f'(Q) - 2 H H' f'(Q) - 4 Q H H' f''(Q) \, .
\end{align}
Here $H'=\frac{dH}{dN}$. 
Eq.~\eqref{vQ_FLRW0phismplN} can be solved with respect to $\omega$ and $V$, again, as follows, 
\begin{align}
\label{vQ_FLRW0phiomgVN}
\omega(\phi) =&\, - \frac{2 H'}{H} f'(Q) + 24 H H' f''(Q) \, , \nonumber \\
V(\phi) =&\, \frac{f(Q)}{2} - Q f'(Q) - HH' f'(Q) - 2 Q H H' f''(Q) \, .
\end{align}
This tells that by using a function $\eta_N(\phi)$, if we consider the model, 
\begin{align}
\label{vQ_FLRW0etaN}
\omega(\phi) =&\, - \frac{2{\eta_N}'(\phi)}{\eta_N(\phi)} f'\left( - 6\eta_N(\phi)^2 \right) + 24 \eta_N(\phi) {\eta_N}'(\phi) f''\left( - 6 \eta_N(\phi)^2 \right) \, , \nonumber \\
V(\phi) =&\, \frac{f\left( - 6 \eta_N(\phi)^2 \right)}{2} + 6 \eta_N(\phi)^2 f'\left( - 6 \eta_N(\phi)^2 \right) - \eta_N(\phi) {\eta_N}'(\phi) f'\left( -6 \eta_N(\phi)^2 \right) \nonumber \\
&\, + 12 \eta_N(\phi)^3 {\eta_N}'(\phi) f''\left( - 6 \eta_N(\phi)^2 \right) \, ,
\end{align}
we obtain an exact solution, again, 
\begin{align}
\label{solN}
\phi=N \, , \quad H=\eta_N(N)\, .
\end{align}
Then if we choose 
\begin{align}
\label{HubbleNewCN}
\eta_N(\phi) = \frac{H_0}{\left(1 + \e^{\beta\left(\phi-N_1\right)}\right)^\gamma} \, ,
\end{align}
the Hubble rate $H$ in \eqref{HubbleNewC} is reproduced. 
By tuning the parameters, we obtain a model satisfying the constraints by ACT etc. 

As a simple example, we may consider the following functional form of $f(Q)$, 
\begin{align}
\label{fQex}
f (Q) = \frac{1}{2\kappa^2} \left( - Q + \alpha Q^2 \right)\, . 
\end{align}
For the model \eqref{Hex1beta}, we find 
\begin{align}
\label{vQ_FLRW0eta00}
\omega(\phi) 
=&\, - \frac{ \frac{2{H_0}^2\e^{\frac{2 H_0}{\alpha} \left( \phi - t_0 \right)}}{1 + \e^{\frac{2 H_0}{\alpha} \left( \phi - t_0 \right)}}}
{\kappa^2 \left\{ 1 + \alpha \ln \left( 1 + \e^{\frac{2 H_0}{\alpha} \left( \phi - t_0 \right)} \right) \right\}^2} 
\left[ -1 + \frac{36\alpha H_0}{\left\{1 + \alpha \ln \left( 1 + \e^{\frac{2 H_0}{\alpha} \left( \phi - t_0 \right)} \right)\right\}^2} \right] \, , \nonumber \\
V(\phi) 
=&\, \frac{1}{2\kappa^2} \left[ - \frac{3{H_0}^2}{\left\{1 + \alpha \ln \left( 1 + \e^{\frac{2 H_0}{\alpha} \left( \phi - t_0 \right)} \right)\right\}^2}
 - \frac{54\alpha {H_0}^4}{\left\{1 + \alpha \ln \left( 1 + \e^{\frac{2 H_0}{\alpha} \left( \phi - t_0 \right)} \right)\right\}^4} \right. \nonumber \\
&\, \left. \qquad - \frac{ \frac{2{H_0}^2\e^{\frac{2 H_0}{\alpha} \left( \phi - t_0 \right)}}{1 + \e^{\frac{2 H_0}{\alpha} \left( \phi - t_0 \right)}}}
{\left\{ 1 + \alpha \ln \left( 1 + \e^{\frac{2 H_0}{\alpha} \left( \phi - t_0 \right)} \right) \right\}^2} 
\left\{ 1 + \frac{36\alpha {H_0}^2}{\left\{1 + \alpha \ln \left( 1 + \e^{\frac{2 H_0}{\alpha} \left( \phi - t_0 \right)} \right)\right\}^2} \right\} \right]\, .
\end{align}
On the other hand, for the model \eqref{HubbleNewCN}, we obtain, 
\begin{align}
\label{vQ_FLRW0etaN00}
\omega(\phi) 
=&\, - \frac{\beta\gamma H_0 \e^{\beta\left(\phi-N_1\right)}}{\kappa^2 \left(1 + \e^{\beta\left(\phi-N_1\right)}\right)^{\gamma+1}} 
\left[ - \frac{\left(1 + \e^{\beta\left(\phi-N_1\right)}\right)^\gamma}{H_0} + \frac{24 \alpha H_0}{\left(1 + \e^{\beta\left(\phi-N_1\right)}\right)^\gamma} \right] \, , \nonumber \\
V(\phi) 
=&\, \frac{1}{2\kappa^2} \left[ - \frac{3{H_0}^2}{\left(1 + \e^{\beta\left(\phi-N_1\right)}\right)^{2\gamma}} 
 - \frac{54\alpha{H_0}^4}{\left(1 + \e^{\beta\left(\phi-N_1\right)}\right)^{4\gamma}} \right. \nonumber \\
&\, \left. \qquad - \frac{\beta\gamma H_0 \e^{\beta\left(\phi-N_1\right)}}{\left(1 + \e^{\beta\left(\phi-N_1\right)}\right)^{\gamma+1}} 
\left\{ \frac{H_0}{\left(1 + \e^{\beta\left(\phi-N_1\right)}\right)^\gamma} 
+ \frac{36\alpha{H_0}^3}{\left(1 + \e^{\beta\left(\phi-N_1\right)}\right)^{3\gamma}} \right\} \right] \, . 
\end{align}
Thus, we obtained ACT-consistent inflation for the above theory. 
Note that this result was achieved by the corresponding tuning of scalar potentials. 
In other words, the above quadratic gravity without scalar does not lead to ACT inflation. 
Just in the same way, one can construct ACT-consistent inflation for large classes of $f(Q)$ gravity.

\section{The correspondence of generalised entropic cosmology theory with $f(Q)$ modified gravity\label{Sec6}}

Jacobson~\cite{Jacobson:1995ab} has shown that the Einstein equation can be obtained from the Bekenstein-Hawking entropy by thermodynamical considerations. 
Recently, various kinds of generalised entropies have been proposed~\cite{Tsallis:1987eu, Renyi, Barrow:2020tzx, SayahianJahromi:2018irq, Kaniadakis:2005zk, Drepanou:2021jiv, Majhi:2017zao, Liu:2021dvj, Nojiri:2022aof, Nojiri:2022dkr, Odintsov:2022qnn}.
Then, a natural question could be what the gravity theories correspond to generalised entropy? 
In \cite{Nojiri:2025fiu}, we found the correspondence between $f(T)$ and $f(Q)$ gravities and general entropic cosmology. 
Because $f(T)$ and $f(Q)$ gravity theories are local theories, we can discuss the local dynamics, such as the fluctuations and the gravitational waves. 

\subsection{Gravity and thermodynamics\label{Sec6A}}

The relation between gravity and thermodynamics has been well studied after it was found that a black hole has a temperature and entropy, which is the Bekenstein-Hawking entropy~\cite{Bekenstein:1973ur, Hawking:1975vcx}, 
\begin{align}
S = \frac{A}{4G} \, .
\label{BH-entropy}
\end{align}
Here $A = 4\pi r_\mathrm{H}^2$ is the area of the horizon given by the horizon radius $r_\mathrm{H}$, and $G$ is Newton's gravitational constant. 
After that, Jacobson has shown that the Einstein equation can be obtained from the Bekenstein-Hawking entropy by thermodynamical considerations, \cite{Jacobson:1995ab}. 
Jacobson has assumed that the equation can be written in terms of the spacetime curvature. 
Or more exactly, the Raychaudhuri equation used in the derivation includes curvature. 

In different physical and statistical systems, various entropies exist.
The Tsallis entropy could describe the thermodynamics in non-extensive sysytems $\left(S_{A+B}\neq S_A + S_B\right)$ \cite{Tsallis:1987eu}, 
\begin{align}
S_\mathrm{T} = \frac{A_0}{4G}\left(\frac{A}{A_0}\right)^{\delta} \, .
\label{Tsallis entropy}
\end{align}
Here $A_0$ is a constant with the dimension of area, $\delta$ is an exponent. 

The R\'{e}nyi entropy is used in quantum information and entanglement entropy~\cite{Renyi}, 
\begin{align}
S_\mathrm{R} = \frac{1}{\alpha} \ln \left( 1 + \alpha S \right) \, .
\label{Renyi entropy}
\end{align}
Here $\alpha$ is a parameter and $S$ is the Bekenstein-Hawking entropy. 

Barrow considered entropy from the viewpoint of quantum fluctuation \cite{Barrow:2020tzx}, 
\begin{align}
\label{Barrow-entropy}
S_\mathrm{B} = \left(\frac{A}{A_\mathrm{Pl}} \right)^{1+\Delta/2} \, .
\end{align}
The area of the black hole horizon is expressed by $A$, and $A_\mathrm{Pl} = 4G$ is called the Planck area. 

The Sharma-Mittal entropy is a generalisation of both the R{\'{e}}nyi and Tsallis entropy~\cite{SayahianJahromi:2018irq}. 
The original papers by Sharma and Mittal can be found in \cite{SharmaMittal1, SharmaMittal2}. 
The expression of the Sharma-Mittal entropy is given by 
\begin{align}
S_\mathrm{SM} = \frac{1}{R}\left[\left(1 + \delta ~S\right)^{R/\delta} - 1\right] \, , 
\label{SM entropy}
\end{align}
which has two parameters $R$ and $\delta$. 

The Kaniadakis entropy was proposed from the consistency with special relativity \cite{Kaniadakis:2005zk, Drepanou:2021jiv}, 
\begin{align}
S_\mathrm{K} = \frac{1}{K}\sinh{\left(KS\right)} \, .
\label{K-entropy}
\end{align}
A phenomenological parameter is expressed by $K$. 

The Loop Quantum Gravity entropy\cite{Majhi:2017zao, Liu:2021dvj} is motivated from loop quantum gravity, 
\begin{align}
S_q = \frac{1}{\left(1-q\right)}\left[\mathrm{e}^{(1-q)\Lambda(\gamma_0)S} - 1\right] \, .
\label{LQG entropy}
\end{align}
Here $q$ is the exponent and $\Lambda(\gamma_0) = \ln{2}/\left(\sqrt{3}\pi\gamma_0\right)$ is a function of the Barbero-Immirzi parameter $\gamma_0$. 

After that, generalised entropies were proposed \cite{Nojiri:2022aof, Nojiri:2022dkr, Odintsov:2022qnn}. 
These entropies give an interpolation between several generalised entropies, although there is no such strong physical motivation. 
Nevertheless, a corresponding microscopic description of generalised entropy was also developed.
Four ($\alpha_{\pm}$, $\delta$, $\gamma$)- and six ($\alpha_{\pm}$, $\delta_{\pm}$, $\gamma_{\pm}$)-parameter generalised entropies were proposed as \cite{Nojiri:2022aof, Nojiri:2022dkr}, 
\begin{align}
\label{intro-1}
S_4 \left(\alpha_{\pm},\delta,\gamma\right) =&\, \frac{1}{\gamma}\left[\left(1 + \frac{\alpha_+}{\delta} S\right)^{\delta}
 - \left(1 + \frac{\alpha_-}{\delta} S\right)^{-\delta}\right] \, , \\
\label{intro-2}
S_6 \left(\alpha_{\pm},\delta_{\pm},\gamma_{\pm}\right) =&\, \frac{1}{\alpha_+ + \alpha_-}
\left[ \left( 1 + \frac{\alpha_+}{\delta_+} S^{\gamma_+} \right)^{\delta_+} 
 - \left( 1 + \frac{\alpha_-}{\delta_-} S^{\gamma_-} \right)^{-\delta_-} \right]\, .
\end{align}
For some parameter choices, as particular examples, the above two entropies describe all the known entropies. 
A simplified version of the four- and six-parameter generalised entropies is the three-parameter ($\alpha$, $\delta$, $\gamma$) entropy \cite{Nojiri:2022aof}, 
\begin{align}
S_3\left(\alpha,\delta,\gamma\right) = \frac{1}{\gamma}\left[\left(1 + \frac{\alpha}{\delta} S\right)^\delta - 1\right] \, .
\label{intro-3}
\end{align}
$S_3$ gives all the above entropies except the Kaniadakis entropy. 
The five-parameter ($\alpha_{\pm}$, $\delta,\gamma$, $\epsilon$) entropy was proposed to solve the problem of singularity when $H\to 0$ \cite{Odintsov:2022qnn}, 
{\small 
\begin{align}
S_5\left(\alpha_{\pm},\delta,\gamma,\epsilon\right) 
= \frac{1}{\gamma}\left[\left\{1 + \frac{1}{\epsilon}\tanh\left(\frac{\epsilon \alpha_+}{\delta} S\right)\right\}^{\delta}
 - \left\{1 + \frac{1}{\epsilon}\tanh\left(\frac{\epsilon \alpha_-}{\delta} S\right)\right\}^{-\delta} \right] \, .
\label{intro-4}
\end{align}
Then, a natural question is what could be the gravity theories corresponding to generalised entropy? 

\subsection{Generalised entropies and cosmology\label{Sec6B}}

As a relation between the entropy and gravity, a simplified version is the relation between the entropy and cosmology. 
In this subsection, we consider a further simplified version. 
For a more detailed and rigorous version, see \cite{Cai:2005ra, Akbar:2006er}. 

We consider the FLRW spacetime with a flat spatial part \eqref{FLRW}. 
The radius $r_\mathrm{H}$ of the cosmological horizon is given by 
\begin{align}
\label{apphor}
r_\mathrm{H}=\frac{1}{H}\, .
\end{align}
Here, $H = \dot{a}/a$ is the Hubble rate, again. 

The Bekenstein-Hawking entropy as an entropy inside the cosmological horizon is given by, 
\begin{align}
\label{BH-entropy2}
S = \frac{A}{4G} = \frac{4\pi {r_\mathrm{H}}^2}{4G} = \frac{4\pi}{4G H^2} \, .
\end{align}
Because the flux of the energy $E$ should be equal to the decrease of the heat $\mathcal{Q}$ in the region, we find
\begin{align}
\label{Tslls2}
d\mathcal{Q} = - dE = -\frac{4\pi}{3} r_\mathrm{H}^3 \dot\rho dt = -\frac{4\pi}{3H^3} \dot\rho dt 
= \frac{4\pi}{H^2} \left( \rho + p \right) dt \, .
\end{align}
Here $\rho$ and $p$ are the energy density and the pressure of the cosmological fluid, that is, matter, again. 
We have used the conservation law \eqref{cons}. 

The Hawking temperature is given by 
\begin{align}
\label{Tslls6}
T = \frac{1}{2\pi r_\mathrm{H}} = \frac{H}{2\pi}\, .
\end{align}
Then the first law of thermodynamics $TdS = d\mathcal{Q}$ gives $\dot H = - 4\pi G \left( \rho + p \right) = \frac{4\pi G}{3H} \dot \rho$. 
By integrating the equation with respect to the cosmological time $t$, we obtain 
\begin{align}
\label{Tslls8}
H^2 = \frac{8\pi G}{3} \rho + \frac{\Lambda}{3}\, , 
\end{align}
which is nothing but the first Friedmann equation given by the Einstein equation. 
The parameter $\Lambda$ appears as a constant of the integration and is regarded as a cosmological constant. 
Then we find the correspondence between thermodynamics and Einstein's gravity. 

In the case of generalised entropy $S_\mathrm{g}$, the first law of thermodynamics tells, 
\begin{align}
\dot{H}\left. \left(\frac{\partial S_\mathrm{g}}{\partial S}\right) \right|_{S = \frac{A}{4G} = \frac{\pi}{GH^2}}= -4\pi G\left(\rho + p\right) \, .
\label{FRW1-subB}
\end{align}
By using the conservation law \eqref{cons}, we find 
\begin{align}
\label{SgH}
\left. \left(\frac{\partial S_\mathrm{g}}{\partial S}\right) \right|_{S = \frac{A}{4G} = \frac{\pi}{GH^2}} d\left( H^2 \right) = \left(\frac{8\pi G}{3}\right)d\rho \, . 
\end{align}
By integrating the above expression \eqref{SgH}, we obtain a generalised first Friedmann equation, 
\begin{align}
\label{gSeq}
\mathcal{H} \left( H^2 \right)^2 \equiv 
\int^{H^2} dx \left. \left(\frac{\partial S_\mathrm{g}}{\partial S}\right) \right|_{S = \frac{\pi}{Gx}} = \left(\frac{8\pi G}{3}\right)\rho \, . 
\end{align}
In the case of four-parameter generalised entropies \eqref{intro-1}, the equation has the following form, 
\begin{align}
\frac{GH^4\beta}{\pi\gamma}&\,\left[ \frac{1}{\left(2+\beta\right)}\left(\frac{GH^2\beta}{\pi\alpha_-}\right)^{\beta}
{}_2F_{1}\left(1+\beta, 2+\beta; 3+\beta; -\frac{GH^2\beta}{\pi\alpha_-}\right) \right. \nonumber\\ 
&\, \left. + \frac{1}{\left(2-\beta\right)}\left(\frac{GH^2\beta}{\pi\alpha_+}\right)^{-\beta}
{}_2F_{1}\left(1-\beta, 2-\beta; 3-\beta; -\frac{GH^2\beta}{\pi\alpha_+}\right) \right] \nonumber \\
=&\, \frac{8\pi G\rho}{3} + \frac{\Lambda}{3} \, .
\label{FRW-2}
\end{align}
The cosmological constant $\Lambda$ appears as an integration constant, again. 
The function $_2F_1$ is the hypergeometric function. 
\begin{align}
\label{hgf}
{}_2F_{1}\left(\alpha, \beta; \gamma;z\right)= 1 + \frac{\alpha\beta}{1!\gamma} z + \frac{\alpha(\alpha+1)\beta(\beta+1)}{2! \gamma(\gamma+1)} z^2 + \cdots\, .
\end{align}
The other entropies can be obtained in the following limiting procedures. 
\begin{itemize}
\item In the case of the Tsallis entropy~\eqref{Tsallis entropy} or the Barrow entropy~\eqref{Barrow-entropy}, 
by identifying $\beta = \delta$ or $\beta = 1 + \Delta$, respectively, first we choose $\alpha_- \to 0$. 
After that, by choosing $\gamma = \left(\alpha_+/\beta\right)^{\beta}$ and keeping $\beta$ finite, we consider the limit of $\alpha_+ \rightarrow \infty$. 
Then we obtain the Tsallis entropy~\eqref{Tsallis entropy} or the Barrow entropy~\eqref{Barrow-entropy}. 
\item The R\'{e}nyi entropy~\eqref{Renyi entropy} can be obtained in the limit $\alpha_- = 0$, $\beta \rightarrow 0$, and $\alpha \equiv \frac{\alpha_+}{\beta} \rightarrow \mathrm{finite}$ and by choosing $\gamma = \alpha_+$. 
\item The Sharma-Mittal entropy~\eqref{SM entropy} is obtained in the limit of $\alpha_- \to 0$ by identifying $\gamma = R$, $\alpha_+ = R$, and $\beta = R/\delta$.
\item In the limit of $\beta \rightarrow \infty$, by identifying $\alpha_+ = \alpha_- = \frac{\gamma}{2} = K$, we obtain the Kaniadakis entropy~\eqref{K-entropy}. 
\item The Loop Quantum Gravity entropy~\eqref{LQG entropy} with $\Lambda(\gamma_0) = 1$ or equivalently $\gamma_0 = \frac{\ln{2}}{\pi\sqrt{3}}$ is obtained in the limit $\alpha_- = 0$, by identifying $\beta \rightarrow \infty$ and $\gamma = \alpha_+ = (1-q)$.
\end{itemize}

A question is what gravity theories could correspond to the generalised entropies. 
In the case of $F(R)$ gravity, 
\begin{align}
\label{FR}
S = \frac{1}{2\kappa^2} \int d^4 x \sqrt{-g} F(R) + S_\mathrm{matter}\, ,
\end{align}
the equation corresponding to the first Friedmann equation has the following form, 
\begin{align}
\label{FR1st}
\frac{1}{\kappa^2}\left(\frac{1}{2}F(R) - 3\left(H^2 + \dot H\right) F'(R) + 18 \left(4H^2 \dot H + H \ddot H\right)F''(R)\right) = \rho \, .
\end{align}
Because the l.h.s. in \eqref{FR1st} includes $\dot H$, $\ddot H$, but only $H$ appears via the Bekenstein-Hawking entropy as in \eqref{BH-entropy2}, the $F(R)$ gravity cannot correspond to generalised entropies. 

\subsection{Correspondence between $f(Q)$ gravity and generalised entropies\label{Sec6C}}

We now consider the correspondence between $f(Q)$ gravity and generalised entropies. 
The equation corresponding to the first Friedmann equation is given by 
\begin{align}
\label{vQ_FLRW0B}
\frac{1}{3} \left( \frac{f(Q)}{2} + 6 H^2 f'(Q) \right) = \left( \frac{8\pi G}{3} \right)\rho \, . 
\end{align}
In the spatially flat FLRW spacetime~\eqref{FLRW}, $Q=-6H^2$. 
Compared with \eqref{gSeq}, we find 
\begin{align}
\label{vQ_FLRW1B}
\frac{1}{3} \left( \frac{f(Q)}{2} + 6 H^2 f'(Q) \right) = \mathcal{H} \left( H^2 \right)^2 \equiv 
\int^{H^2} dx \left. \left(\frac{\partial S_\mathrm{g}}{\partial S}\right) \right|_{S = \frac{\pi}{Gx}} \, .
\end{align}
The l.h.s. and r.h.s. of Eq.~\eqref{vQ_FLRW1} are given by a function of only $H$ and therefore we find 
\begin{align}
\label{gSeqFQ2}
f(Q) = 3 \left( - Q \right)^\frac{1}{2} \int^{-\frac{Q}{6}} dy \left( 6 y \right)^{-\frac{3}{2}} 
\int^y dx \left. \left(\frac{\partial S_\mathrm{g}}{\partial S}\right) \right|_{S = \frac{\pi}{Gx}} \, . 
\end{align}
Then we obtain the $f(Q)$ gravity theory corresponding to the generalised entropy $S_\mathrm{g}$. 

Conversely, we may express the generalised entropy in terms of $f(Q)$. 
By solving \eqref{gSeqFQ2} with respect to $S_\mathrm{g}$, we obtain the following expression, 
\begin{align}
\label{Gf2}
S_\mathrm{g} = - \frac{36\pi}{G} \left. \int^Q dq \left( \frac{2 f''(q)}{q} + \frac{f'(q)}{q^2} \right) \right|_{Q= - \frac{6\pi}{GS}} \, .
\end{align}
Especially for the model \eqref{arbQinf} compatible with the ACT observation, because
\begin{align}
\label{arbQinfFF00}
\frac{2 f''(q)}{q} + \frac{f'(q)}{q^2} =
\frac{1}{16\pi G} \left\{ - \frac{2}{q^2} - \frac{4}{q^2\left(1 - \frac{q}{6 {H_0}^2}\right)} - \frac{2}{3{H_0}^2 q \left( 1 - \frac{q}{6 {H_0}^2} \right)^2} \right\} \, ,
\end{align}
we obtain. 
\begin{align}
\label{Gf3}
S_\mathrm{g} =&\, 36 S - \frac{24\pi S}{{H_0}^2 G S + \pi} + S_0 \, .
\end{align}
Here ${\tilde S}_0$ and $S_0$ are constant of the integration and $S_0=- \frac{36\pi {\tilde S}_0}{G}$. 
On the other hand, for the model \eqref{arbCQ}, we obtain
\begin{align}
\label{Gf2C}
S_\mathrm{g} 
=&\, \frac{72\pi\gamma}{G} \left[ 
 - \frac{\rho_0 \e^{4 \left(N_1-N_0\right)}}{\left( 6{H_0}^2 \right)^2} \left( {2 - \frac{4}{\beta\gamma}} \right) 
B_{\left(\frac{\pi}{{H_0}^2GS}\right)^\frac{1}{2\gamma}} \left( - \frac{4}{\beta}, \frac{4}{\beta}+1 \right) 
+ \frac{4\rho_0 \e^{4 \left(N_1-N_0\right)}}{\left(6{H_0}^2\right)^2\beta\gamma} 
B_{\left(\frac{\pi}{{H_0}^2GS}\right)^\frac{1}{2\gamma}} \left( 1 - \frac{4}{\beta}, \frac{4}{\beta} \right) \right] \, .
\end{align}
Therefore, we found an ACT-consistent inflationary cosmology based on the generalised entropy. 
Note that eventually corresponding generalised entropies may give some approach for the fundamental entropy of the primordial universe. 
It is also interesting that even if $f(Q)$ gravity may not give a realistic description of the universe, the above equivalence provides an excellent geometric $f(Q)$ description for entropic cosmology.

Let us now discuss the reheating era. 
In the $f(Q)$ model \eqref{arbQinf} corresponding to the general entropy \eqref{Gf3}, the particle creation by reheating is generated by the source term $J$ in \eqref{conseff} corresponding to $\mathcal{J}_p$ in \eqref{fp}, but the origin of $\mathcal{J}_p$ has not been clarified. 
In the $f(Q)$ model \eqref{arbCQ} corresponding to the general entropy \eqref{Gf2C}, the particle creation has been neglected. 
When we discuss the reheating, a key point is the conservation law \eqref{cons}. 
In the case of an inflaton model, the energy of the matter itself is not conserved by the particle creation. 
The energy is, of course, supplied from the inflaton field, and the total energy combining matter and inflaton is conserved. 
Even in the entropic cosmology, by considering the effective fluid generating the inflation, we may like to make the effective fluid play the role of an instanton field and make the effective fluid supply the energy for the particle creation. 
In this case, the energy of the effective fluid is not conserved because it is converted to the energy of the created particles. 
We should note, however, that when we derive the Friedmann equations from the generalised entropy by the thermodynamical consideration, we have used the conservation law in \eqref{SgH} or in \eqref{Tslls2} in the case of Einstein's gravity. 
The situation has not changed even in the detailed and more rigorous version, see \cite{Cai:2005ra, Akbar:2006er}. 
Therefore, it does not look like the generalised entropy could play the role of the inflaton. 
In this sense, coupling with scalar may give the solution for the correct preheating era.

A possibility is also the quantum effects in the curved spacetime \cite{Parker:1968mv, Parker:1969au, Parker:1971pt, Birrell:1982ix, Buchbinder:1992gdx}, that is, the particle creation in the expanding universe. 
The energy density of the particles created at the end of the inflation can be estimated to be 
\begin{align}
\label{pcq}
\rho \sim {H_\mathrm{end}}^4 \, .
\end{align}
Here $H_\mathrm{end}$ is the value of the Hubble rate $H$ when the inflation ended. 
On the other hand, by the standard inflaton decay, the corresponding energy density of the created particle is 
\begin{align}
\label{infcr}
\rho \sim {M_\mathrm{Planck}}^2 {H_\mathrm{end}}^2\, .
\end{align}
Here $M_\mathrm{Planck}$ is the Planck mass or the Planck energy. 
Because we usually have ${M_\mathrm{Planck}}^2 \gg {H_\mathrm{end}}^2$, the particles created by the quantum effect can be negligible. 
If there is no inflaton, as in the case of the $f(Q)$ gravity or the entropic cosmology based on the generalised entropy, the particle creation by the quantum effects becomes leading. 
We should note that the quantum particle creation is consistent with energy conservation because the energy is supplied from the change of the vacuum. 

\subsection{The properties of the obtained entropies}

In \cite{Nojiri:2022aof, Nojiri:2022dkr}, the following properties are required for the generalised entropy. 

\begin{enumerate}

\item {\it Generalised third law}: All these entropies vanish when the Bekenstein-Hawking entropy vanishes. 
In the third law of standard thermodynamics for closed systems in thermodynamic equilibrium, the quantity $\e^S$ expresses the number of states, or the volume of these states, and therefore the entropy $S$ vanishes when the temperature does because the ground (vacuum) state should be unique. 
By contrast, the Bekenstein-Hawking entropy $\mathcal{S}$ diverges when the temperature $T$ vanishes, and it goes to zero at infinite temperature. 
However, requiring any generalised entropy to vanish when the Bekenstein-Hawking entropy $\mathcal{S}$ vanishes could be a natural requirement.

\item {\it Monotonically increasing functions}: All the above entropies are monotonically increasing functions of the Bekenstein-Hawking entropy $S$.

\item {\it Positivity}: All the above entropies are positive, as is the Bekenstein-Hawking entropy~\eqref{BH-entropy2}. 
This is natural because $\e^S$ corresponds to the number of states (or to the volume of these states), which is greater than unity.

\item {\it Bekenstein-Hawking limit}: All the above entropies reduce to the Bekenstein-Hawking entropy~\eqref{BH-entropy2} in an appropriate limit.

\end{enumerate}
All the generalised entropies in Subsection~\ref{Sec6A} satisfy the above conditions. 

By using \eqref{Gf2}, we investigate the properties of $f(Q)$ which satisfy the above conditions. 
Because $Q= - \frac{6\pi}{GS}$, the limit $S\to 0$ corresponds to the limit $Q\to - \infty$. 
Then if the integration $\int^Q dq \left( \frac{2 f''(q)}{q} + \frac{f'(q)}{q^2} \right)$ in \eqref{Gf2} is finite when $Q\to -\infty$, by adjusting the constant of the integration, the {\it Generalised third law} can be satidfied. 
\begin{align}
\label{G3rd}
\mbox{{\it Generalised third law}:}\ \left| \int^{-\infty} dq \left( \frac{2 f''(q)}{q} + \frac{f'(q)}{q^2} \right) \right|<\infty\, .
\end{align}
The relation $Q= - \frac{6\pi}{GS}$ tells that if $S$ increases, $Q$ also increases. 
Then the condition of {\it Monotonically increasing functions} can be satisfied if the integrand $\frac{2 f''(q)}{q} + \frac{f'(q)}{q^2}$ is always negative. 
If the conditions of {\it Generalised third law} and {\it Monotonically increasing functions} are satisfied, the condition of {\it Positivity} is always satisfied. 
\begin{align}
\label{MIF} 
\mbox{{\it Monotonically increasing functions} and {\it Positivity}:}\ \frac{2 f''(Q)}{Q} + \frac{f'(Q)}{Q^2}<0\, .
\end{align}
The condition of the {\it Bekenstein-Hawking limit} requires, 
\begin{align}
\label{BH1}
\left. \frac{\partial S_\mathrm{g}}{\partial S}\right|_{S\to 0} = 1\, .
\end{align}
Because 
\begin{align}
\label{SQrltn}
\frac{\partial}{\partial S} = \frac{dQ}{dS}\frac{\partial}{\partial Q} = \frac{GQ^2}{6\pi}\frac{\partial}{\partial Q} \, ,
\end{align}
we find that the condition of the {\it Bekenstein-Hawking limit} is satisfied if the following is satisfied, 
\begin{align}
\label{BH2}
\mbox{{\it Bekenstein-Hawking limit}:}\ 
\left. \left[ - 6Q^2 \left( \frac{2 f''(Q)}{Q} + \frac{f'(Q)}{Q^2} \right) \right] \right|_{Q\to -\infty} 
= \left. \left[ - 12 Q f''(Q) - 6 f'(Q) \right] \right|_{Q\to -\infty} =1\, .
\end{align}
For the model \eqref{arbQinfFF}, we find 
\begin{align}
\label{arbQinfFFfdfdd}
\frac{2 f''(q)}{q} + \frac{f'(q)}{q^2} =&\, 
\frac{1}{16\pi G} \left\{ - \frac{2}{q^2} 
 - \frac{4}{q^2\left(1 - \frac{q}{6 {H_0}^2}\right)}
 - \frac{2}{3{H_0}^2 q \left( 1 - \frac{q}{6 {H_0}^2} \right)^2} \right\} \, .
\end{align}
And for the model of $f(Q)$ gravity in \eqref{arbCQB}, we find 
\begin{align}
\label{arbCQB00fdfdd}
\frac{2 f''(q)}{q} + \frac{f'(q)}{q^2} =&\, 
 - \frac{\rho_0 \e^{4 \left(N_1-N_0\right)}}{\left( 6{H_0}^2 \right)^3} \left( {2 - \frac{4}{\beta\gamma}} \right) \left( -\frac{q}{6{H_0}^2} \right)^{- \frac{2}{\beta\gamma}-1}
\left( 1 - \left(-\frac{q}{6{H_0}^2}\right)^\frac{1}{2\gamma} \right)^\frac{4}{\beta} \nonumber \\
&\, + \frac{4\rho_0 \e^{4 \left(N_1-N_0\right)}}{\left(6{H_0}^2\right)^3\beta\gamma} \left( -\frac{q}{6{H_0}^2} \right)^{\frac{1}{2\gamma} - \frac{2}{\beta\gamma}-1} 
\left( 1 - \left(-\frac{q}{6{H_0}^2}\right)^\frac{1}{2\gamma} \right)^{\frac{4}{\beta} -1} \, .
\end{align}
We should note that there are singularities at $q=-6{H_0}^2$ for both of the expressions in \eqref{arbQinfFFfdfdd} and \eqref{arbCQB00fdfdd}. 
Therefore, we cannot extend the expressions for large $q$ and the conditions of {\it Generalised third law} and {\it Bekenstein-Hawking limit} are not satisfied for the corresponding entropies \eqref{Gf3} and \eqref{Gf2C} as they are. 
When we consider inflation, we limit the region where $H$ is larger than the Hubble rate at the end of inflation, given by $\epsilon=1$ as in \eqref{slowroll2} and \eqref{endN}, and $H$ is equal to or smaller than the Hubble rate $H_0$ during inflation. 
We may always, however, extrapolate $H$ so that the entropy satisfies the conditions. 
For examples, by modifying \eqref{arbQinfFFfdfdd} and \eqref{arbCQB00fdfdd} as follows, 
\begin{align}
\label{arbQinfFFfdfddm}
\frac{2 f''(q)}{q} + \frac{f'(q)}{q^2} =&\, 
\frac{1}{16\pi G} \left\{ - \frac{2}{q^2} 
 - \frac{4}{q^2\left(1 - \frac{q}{6 {H_0}^2}\right)}
 - \frac{2}{3{H_0}^2 q \left( 1 - \frac{q}{6 {H_0}^2} \right)^2} \right\} \nonumber \\
\to &\, \frac{1}{16\pi G} \left\{ - \frac{2}{q^2} 
 - \frac{4}{q^2\sqrt{\left(1 - \frac{q}{6 {H_0}^2}\right)^2 + {\varepsilon_0}^2}}
 - \frac{2}{3{H_0}^2 q \left\{\left( 1 - \frac{q}{6 {H_0}^2} \right)^2+ {\varepsilon_0}^2 \right\}} \right\} \, , \\
\label{arbCQB00fdfddm}
\frac{2 f''(q)}{q} + \frac{f'(q)}{q^2} =&\, 
 - \frac{\rho_0 \e^{4 \left(N_1-N_0\right)}}{\left( 6{H_0}^2 \right)^3} \left( {2 - \frac{4}{\beta\gamma}} \right) \left( -\frac{q}{6{H_0}^2} \right)^{- \frac{2}{\beta\gamma}-1}
\left( 1 - \left(-\frac{q}{6{H_0}^2}\right)^\frac{1}{2\gamma} \right)^\frac{4}{\beta} \nonumber \\
&\, + \frac{4\rho_0 \e^{4 \left(N_1-N_0\right)}}{\left(6{H_0}^2\right)^3\beta\gamma} \left( -\frac{q}{6{H_0}^2} \right)^{\frac{1}{2\gamma} - \frac{2}{\beta\gamma}-1} 
\left( 1 - \left(-\frac{q}{6{H_0}^2}\right)^\frac{1}{2\gamma} \right)^{\frac{4}{\beta} -1} \nonumber \\
\to &\, - \frac{\rho_0 \e^{4 \left(N_1-N_0\right)}}{\left( 6{H_0}^2 \right)^3} \left( {2 - \frac{4}{\beta\gamma}} \right) \left( -\frac{q}{6{H_0}^2} \right)^{- \frac{2}{\beta\gamma}-1}
\left( 1 - \left( \left( -\frac{q}{6{H_0}^2}\right)^2 + {\varepsilon_0}^2 \right)^\frac{1}{4\gamma}\right)^\frac{4}{\beta} \nonumber \\
&\, + \frac{4\rho_0 \e^{4 \left(N_1-N_0\right)}}{\left(6{H_0}^2\right)^3\beta\gamma} \left( -\frac{q}{6{H_0}^2} \right)^{\frac{1}{2\gamma} - \frac{2}{\beta\gamma}-1} 
\left( 1 - \left( \left( -\frac{q}{6{H_0}^2}\right)^2 + {\varepsilon_0}^2 \right)^\frac{1}{4\gamma}\right)^{\frac{4}{\beta} - 1} \, .
\end{align}
Here $\varepsilon_0$ is a small constant. 
Then the singularity at $q=-6{H_0}^2$ can be avoided, and we may consider the region $q<-6{H_0}^2$ or $H>H_0$. 
Furthermore, we may add more terms that are only relevant for large $H$ so that the conditions could be satisfied. 
The limit that the Bekenstein-Hawking entropy goes to zero corresponds to the limit that $H$ goes to infinity beyond the inflation, which might be a little bit unphysical and quantum gravity effects could be relevant there. 
Therefore, we may not need to consider the region where $H$ is large, $H>H_0$. 

We now consider the reason for the singularities at $q=-6{H_0}^2$ for \eqref{arbQinfFFfdfdd} and \eqref{arbCQB00fdfdd}. 
We rewrite \eqref{gSeq} as follows, 
\begin{align}
\label{gSeq0}
\mathcal{H} \left( H^2 \right)^2 = - \frac{\pi}{G} \int^S \frac{dS}{S^2} \frac{\partial S_\mathrm{g}}{\partial S} = \left(\frac{8\pi G}{3}\right)\rho \, .
\end{align}
During the inflation, $S= \frac{\pi}{G H^2}$ is almost constant, but the energy density $\rho$ and therefore $\mathcal{H} \left( H^2 \right)^2$ rapidly decrease during the inflation due to the expansion. 
Therefore, during the inflation, $\frac{dS}{d\left( \mathcal{H}^2 \right)}$ is very small or $\frac{d\left( \mathcal{H}^2 \right)}{dS}$ is very large. 
Eq.~\eqref{gSeq0} tells, $\frac{1}{S^2} \frac{\partial S_\mathrm{g}}{\partial S}$ and therefore $\frac{\partial S_\mathrm{g}}{\partial S}$ becomes very large when $S=S_0 = \frac{\pi}{G {H_0}^2}$. 
Here $H_0$ is the value of the Hubble rate during the inflation. 
The divergences of \eqref{arbQinfFFfdfdd} and \eqref{arbCQB00fdfdd} occur for this reason. 
Physically, $\frac{\partial S_\mathrm{g}}{\partial S}$ need not diverge, but it can only become very large when $S=S_0 = \frac{\pi}{G {H_0}^2}$. 
For the generalised entropies in Subsection~\ref{Sec6A}, however, it is difficult to make them have this property as long as the parameters in the entropies are positive. 
Then we consider a new type of generalised entropy. 

We now propose the following simple generalised entropy, 
\begin{align}
\label{newS}
S_\mathrm{g} = \left( 1 - \sigma_0 \right) S + \frac{\sigma_0 S}{1 + \frac{S^2}{{S_0}^2}} \, , \quad 0<\sigma_0<1 \, .
\end{align}
Here $S_0$ and $\sigma_0$ are constants. 
As long as $\sigma_0$ satisfies the condition $0<\sigma_0<1$, $S_\mathrm{g}$ in \eqref{newS} is positive and therefore the condition of {\it Positivity} is satisfied. 
We can also easily find that when $S\to 0$, $S_\mathrm{g} \to S$. 
Therefore, the conditions of {\it Generalised third law} and {\it Bekenstein-Hawking limit} are satisfied. 
Because 
\begin{align}
\label{mif}
\frac{\partial S_\mathrm{g}}{\partial S} =1 - \sigma_0 + \frac{\sigma_0}{1 + \frac{S^2}{{S_0}^2}} + \frac{2\sigma_0 S^2 }{\left(1 + \frac{S^2}{{S_0}^2}\right)^2} >0 \, ,
\end{align}
we find the condition of {\it Monotonically increasing functions} is also satisfied. 
Therefore, all the conditions for generalised entropies are satisfied in the entropy \eqref{newS}. 

For the entropy \eqref{newS}. by using \eqref{gSeq0}, we find 
\begin{align}
\label{calH}
\mathcal{H}^2 = \frac{\pi}{GS} \left( 1 - \frac{\frac{\sigma_0 S^2}{{S_0}^2}}{1 + \frac{S^2}{{S_0}^2}} \right) \, .
\end{align}
Because 
\begin{align}
\label{calHd}
\frac{d \left( \mathcal{H}^2 \right)}{d S}
= - \frac{\pi}{G} \left\{ \frac{1}{S^2} \left( 1 - \sigma_0 + \frac{\sigma_0}{1 + \frac{S^2}{{S_0}^2}} \right) + \frac{2\sigma_0}{\left(1 + \frac{S^2}{{S_0}^2}\right)^2} \right\} \, ,
\end{align}
we obtain, 
\begin{align}
\label{calHdS_0}
\left. \frac{d \left( \mathcal{H}^2 \right)}{d S} \right|_{S=S_0}
= - \frac{\pi}{G} \left\{ \frac{1}{{S_0}^2} \left( 1 - \frac{\sigma_0}{2} \right) + \frac{\sigma_0}{2} \right\} \, .
\end{align}
If $S_0\to 0$, $\left. \frac{d \left( \mathcal{H}^2 \right)}{d S} \right|_{S=S_0}$ becomes singular. 
During the inflation, we have $S\sim S_0$. 
Because $\mathcal{H}^2 \to \frac{\pi}{GS}$ when $0<S\ll S_0$ and $\mathcal{H}^2 \to \frac{\pi\left( 1 - \sigma \right)}{GS}$ when $S\gg S_0$, in order to obtain sufficient $e$-folding number during the inflation, we need to require
\begin{align}
0<1 - \sigma \ll 1 \, .
\end{align}
Eq.~\eqref{calH} can be solved with respect to $S$, 
\begin{align}
\label{ScalH}
S =&\, \alpha_+ + \alpha_- - \frac{\pi \left( 1 - \sigma \right) }{3G \mathcal{H}^2} \, , \nonumber \\
{\alpha_\pm}^3 \equiv&\, \frac{1}{2} \left\{ \frac{4\pi^3 \left( 1 - \sigma \right)^3 }{27G^3 \mathcal{H}^6} + \frac{\pi \left( 4 - \sigma \right) {S_0}^2}{3G \mathcal{H}^2} \right\} 
\pm \frac{1}{2} \sqrt{D} \, ,\nonumber \\
D \equiv&\, \left\{ \frac{4\pi^3 \left( 1 - \sigma \right)^3 }{27G^3 \mathcal{H}^6} + \frac{\pi \left( 4 - \sigma \right) {S_0}^2}{3G \mathcal{H}^2} \right\}^2 
+ \frac{1}{27} \left\{ \frac{\pi^2 \left( 1 - \sigma \right)^2 }{G^2 \mathcal{H}^4} + {S_0}^2 \right\}^3 \, .
\end{align}
We should note that $D$ is always positive, $D>0$. 
Then we also find the expression of the Hubble rate $H$ as follows, 
\begin{align}
\label{HcalH}
H^2 = \frac{\pi}{GS} = \frac{\pi}{G \left( \alpha_+ + \alpha_- - \frac{\pi \left( 1 - \sigma \right) }{3G \mathcal{H}^2} \right)} \, .
\end{align}
In order to find the $e$-folding number $N$ dependence of $H$, we now assume that there is only radiation. 
Then the energy density is given by $\rho = {\tilde\rho}_0 a^{-4} = \rho_0 \e^{-4N}$, where the origin of $N$ is absorbed into $\rho_0$. 
Because Eq.~\eqref{gSeq0} tells 
\begin{align}
\label{gSeq000}
\mathcal{H}^2 = \frac{8\pi G \rho_0 \e^{-4N}}{3} \, ,
\end{align}
we obtain the $N$ dependence of $\alpha_\pm$ and $D$, 
\begin{align}
\label{alphaD}
{\alpha_\pm}^3 
=&\, \frac{1}{2} \left\{ \frac{\left( 1 - \sigma \right)^3 \e^{12N}}{128G^6 {\rho_0}^3} + \frac{\left( 4 - \sigma \right) {S_0}^2\e^{4N}}{4 G^2 \rho_0} \right\} \pm \frac{1}{2} \sqrt{D} \, , 
\nonumber \\
D 
=&\, \left\{ \frac{\left( 1 - \sigma \right)^3 \e^{12N}}{128G^6 {\rho_0}^3} + \frac{\left( 4 - \sigma \right) {S_0}^2\e^{4N}}{4 G^2 \rho_0} \right\}^2 
+ \frac{1}{27} \left\{ \frac{9\left( 1 - \sigma \right)^2 \e^{8N}}{64 G^4 {\rho_0}^2} + {S_0}^2 \right\}^3 \, .
\end{align}
By using \eqref{HcalH}, we find the $N$ dependence of $H$. 
Then the entropic cosmology is given. 
We may further tune the parameters so that the model could satisfy the constraints coming from the ACT observation, etc., but because the functional form of $H$ is very complicated, it is difficult to find the explicit values of the parameters. 

\subsection{Entropic cosmology based on generalised entropies with scalar}

Let us consider a theory where a scalar field couples with the entropic cosmologies based on the generalised entropy, as in Section~\ref{SecVI}. 

For four ($\alpha_{\pm}$, $\delta$, $\gamma$) parameter generalised entropy in \eqref{intro-1}, 
\begin{align}
\label{intro-1_0}
S_4 \left(\alpha_{\pm},\delta,\gamma\right) =&\, \frac{1}{\gamma}\left[\left(1 + \frac{\alpha_+}{\delta} S\right)^{\delta}
 - \left(1 + \frac{\alpha_-}{\delta} S\right)^{-\delta}\right] \, , 
\end{align}
the equation of the entropic cosmology corresponding to the first Friedmann equation is given by \eqref{FRW-2}, 
\begin{align}
\label{FRW-2_0}
\frac{GH^4\beta}{\pi\gamma}&\,\left[ \frac{1}{\left(2+\beta\right)}\left(\frac{GH^2\beta}{\pi\alpha_-}\right)^{\beta}
{}_2F_{1}\left(1+\beta, 2+\beta; 3+\beta; -\frac{GH^2\beta}{\pi\alpha_-}\right) \right. \nonumber\\ 
&\, \left. + \frac{1}{\left(2-\beta\right)}\left(\frac{GH^2\beta}{\pi\alpha_+}\right)^{-\beta}
{}_2F_{1}\left(1-\beta, 2-\beta; 3-\beta; -\frac{GH^2\beta}{\pi\alpha_+}\right) \right] \nonumber \\
=&\, \frac{8\pi G\rho}{3} \, .
\end{align}
Here we dropped the cosmological constant compared with \eqref{FRW-2}. 
By using the conservation law in \eqref{cons}, 
\begin{align}
\label{cons_0}
\dot\rho + 3 H \left( \rho + p \right) = 0 \, ,
\end{align}
we find the equation corresponding to the second Friedmann equation, 
\begin{align}
\label{2Frdmn}
\frac{8\pi G p}{3} =&\, - \frac{8\pi G\rho}{3} - \frac{8\pi G\dot\rho}{9H} \nonumber \\
=&\, - \frac{GH^4\beta}{\pi\gamma} \left[ \frac{1}{\left(2+\beta\right)}\left(\frac{GH^2\beta}{\pi\alpha_-}\right)^{\beta}
{}_2F_{1}\left(1+\beta, 2+\beta; 3+\beta; -\frac{GH^2\beta}{\pi\alpha_-}\right) \right. \nonumber\\ 
&\, \left. + \frac{1}{\left(2-\beta\right)}\left(\frac{GH^2\beta}{\pi\alpha_+}\right)^{-\beta}
{}_2F_{1}\left(1-\beta, 2-\beta; 3-\beta; -\frac{GH^2\beta}{\pi\alpha_+}\right) \right] \nonumber \\
&\, - \frac{GH^2\beta}{\pi\gamma} \left[ \frac{2 \beta \dot H}{3\left(2+\beta\right)} \left(\frac{GH^2\beta}{\pi\alpha_-}\right)^\beta
{}_2F_{1}\left(1+\beta, 2+\beta; 3+\beta; -\frac{GH^2\beta}{\pi\alpha_-}\right) \right. \nonumber \\ 
&\, \left. - \frac{2\beta \dot H}{3\left(2-\beta\right)}
\left(\frac{GH^2\beta}{\pi\alpha_+}\right)^{-\beta}
{}_2F_{1}\left(1-\beta, 2-\beta; 3-\beta; -\frac{GH^2\beta}{\pi\alpha_+}\right) \right] \nonumber \\
&\, + \frac{2GH^2 \dot H\beta}{3\pi\gamma} \left[ \frac{1+\beta}{3+\beta} \left(\frac{GH^2\beta}{\pi\alpha_-}\right)^{\beta+1}
{}_2F_{1}\left(2+\beta, 3+\beta; 4+\beta; -\frac{GH^2\beta}{\pi\alpha_-}\right) \right. \nonumber\\ 
&\, \left. + \frac{1-\beta}{3-\beta} \left(\frac{GH^2\beta}{\pi\alpha_+}\right)^{-\beta+1}
{}_2F_{1}\left(2-\beta, 3-\beta; 4-\beta; -\frac{GH^2\beta}{\pi\alpha_+}\right) \right] \, . 
\end{align}
Here, we have used the formula for Gauss' hypergeometric function, 
\begin{align}
\label{Gauss}
\frac{d}{dz}{}_2F_1(a,b;c;z) = \frac{ab}{c}\,{}_2F_1(a+1,b+1;c+1;z) \, .
\end{align}

We now consider the scalar field as in \eqref{fQactn}, 
\begin{align}
\label{fQactnEN}
S_\phi =\int d^4 x \sqrt{-g} \left( - \frac{1}{2}\omega(\phi) g^{\mu\nu} \partial_\mu \phi \partial_\nu \phi - V(\phi) \right) \, .
\end{align}
Here, we only wrote the part for the scalar field $\phi$ in the action. 

When we identify $\phi$ with the cosmological time $t$, the energy density $\rho$ and the pressure $p$ are given by, 
\begin{align}
\label{rhpph}
\rho=\frac{1}{2}\omega(t) + V(t)\, , \quad p=\frac{1}{2}\omega(t) - V(t)\, .
\end{align}
By using \eqref{FRW-2_0} and \eqref{2Frdmn}, we find 
\begin{align}
\label{enomgV}
\omega(t) =&\, - \frac{3H^2\beta}{8\pi^2\gamma} \left[ \frac{2 \beta \dot H}{3\left(2+\beta\right)} \left(\frac{GH^2\beta}{\pi\alpha_-}\right)^\beta
{}_2F_{1}\left(1+\beta, 2+\beta; 3+\beta; -\frac{GH^2\beta}{\pi\alpha_-}\right) \right. \nonumber \\ 
&\, \left. - \frac{2\beta \dot H}{3\left(2-\beta\right)}
\left(\frac{GH^2\beta}{\pi\alpha_+}\right)^{-\beta}
{}_2F_{1}\left(1-\beta, 2-\beta; 3-\beta; -\frac{GH^2\beta}{\pi\alpha_+}\right) \right] \nonumber \\
&\, + \frac{H^2 \dot H\beta}{4\pi^2\gamma} \left[ \frac{1+\beta}{3+\beta} \left(\frac{GH^2\beta}{\pi\alpha_-}\right)^{\beta+1}
{}_2F_{1}\left(2+\beta, 3+\beta; 4+\beta; -\frac{GH^2\beta}{\pi\alpha_-}\right) \right. \nonumber\\ 
&\, \left. + \frac{1-\beta}{3-\beta} \left(\frac{GH^2\beta}{\pi\alpha_+}\right)^{-\beta+1}
{}_2F_{1}\left(2-\beta, 3-\beta; 4-\beta; -\frac{GH^2\beta}{\pi\alpha_+}\right) \right] \, , \nonumber \\
V(t) =&\, \frac{3H^4\beta}{8\pi^2\gamma} \left[ \frac{1}{\left(2+\beta\right)}\left(\frac{GH^2\beta}{\pi\alpha_-}\right)^{\beta}
{}_2F_{1}\left(1+\beta, 2+\beta; 3+\beta; -\frac{GH^2\beta}{\pi\alpha_-}\right) \right. \nonumber\\ 
&\, \left. + \frac{1}{\left(2-\beta\right)}\left(\frac{GH^2\beta}{\pi\alpha_+}\right)^{-\beta}
{}_2F_{1}\left(1-\beta, 2-\beta; 3-\beta; -\frac{GH^2\beta}{\pi\alpha_+}\right) \right] \nonumber \\
&\, + \frac{3H^2\beta}{16\pi^2\gamma} \left[ \frac{2 \beta \dot H}{3\left(2+\beta\right)} \left(\frac{GH^2\beta}{\pi\alpha_-}\right)^\beta
{}_2F_{1}\left(1+\beta, 2+\beta; 3+\beta; -\frac{GH^2\beta}{\pi\alpha_-}\right) \right. \nonumber \\ 
&\, \left. - \frac{2\beta \dot H}{3\left(2-\beta\right)}
\left(\frac{GH^2\beta}{\pi\alpha_+}\right)^{-\beta}
{}_2F_{1}\left(1-\beta, 2-\beta; 3-\beta; -\frac{GH^2\beta}{\pi\alpha_+}\right) \right] \nonumber \\
&\, - \frac{H^2 \dot H\beta}{8\pi^2\gamma} \left[ \frac{1+\beta}{3+\beta} \left(\frac{GH^2\beta}{\pi\alpha_-}\right)^{\beta+1}
{}_2F_{1}\left(2+\beta, 3+\beta; 4+\beta; -\frac{GH^2\beta}{\pi\alpha_-}\right) \right. \nonumber\\ 
&\, \left. + \frac{1-\beta}{3-\beta} \left(\frac{GH^2\beta}{\pi\alpha_+}\right)^{-\beta+1}
{}_2F_{1}\left(2-\beta, 3-\beta; 4-\beta; -\frac{GH^2\beta}{\pi\alpha_+}\right) \right] \, . 
\end{align}
By replacing $H$ by a function $\eta(\phi)$ in \eqref{enomgV}, $\omega$ and $V$ are expressed as functions of $\phi$. 
In the obtained model, a solution is given by $\phi=t$ and $H=\eta(t)$ as in \eqref{sol}. 
If we choose $\eta(\phi)$ as in \eqref{Hex1beta}, we obtain a model consistent with the ACT observations, again. 

In the case that we identify $\phi$ with the $e$-holding number $N$, instead of \eqref{rhpph}, we obtain 
\begin{align}
\label{rhpphN}
\rho=\frac{1}{2}\omega(N) H^2 + V(N)\, , \quad p=\frac{1}{2}\omega(N) H^2 - V(N)\, .
\end{align}
Eqs.~\eqref{FRW-2_0} and \eqref{2Frdmn} tell, 
\begin{align}
\label{enomgVN}
\omega(N) =&\, - \frac{3\beta}{8\pi^2\gamma} \left[ \frac{2 \beta HH'}{3\left(2+\beta\right)} \left(\frac{GH^2\beta}{\pi\alpha_-}\right)^\beta
{}_2F_{1}\left(1+\beta, 2+\beta; 3+\beta; -\frac{GH^2\beta}{\pi\alpha_-}\right) \right. \nonumber \\ 
&\, \left. - \frac{2\beta HH'}{3\left(2-\beta\right)}
\left(\frac{GH^2\beta}{\pi\alpha_+}\right)^{-\beta}
{}_2F_{1}\left(1-\beta, 2-\beta; 3-\beta; -\frac{GH^2\beta}{\pi\alpha_+}\right) \right] \nonumber \\
&\, + \frac{H H' \beta}{4\pi^2\gamma} \left[ \frac{1+\beta}{3+\beta} \left(\frac{GH^2\beta}{\pi\alpha_-}\right)^{\beta+1}
{}_2F_{1}\left(2+\beta, 3+\beta; 4+\beta; -\frac{GH^2\beta}{\pi\alpha_-}\right) \right. \nonumber\\ 
&\, \left. + \frac{1-\beta}{3-\beta} \left(\frac{GH^2\beta}{\pi\alpha_+}\right)^{-\beta+1}
{}_2F_{1}\left(2-\beta, 3-\beta; 4-\beta; -\frac{GH^2\beta}{\pi\alpha_+}\right) \right] \, , \nonumber \\
V(\phi) =&\, \frac{3H^4\beta}{8\pi^2\gamma} \left[ \frac{1}{\left(2+\beta\right)}\left(\frac{GH^2\beta}{\pi\alpha_-}\right)^{\beta}
{}_2F_{1}\left(1+\beta, 2+\beta; 3+\beta; -\frac{GH^2\beta}{\pi\alpha_-}\right) \right. \nonumber\\ 
&\, \left. + \frac{1}{\left(2-\beta\right)}\left(\frac{GH^2\beta}{\pi\alpha_+}\right)^{-\beta}
{}_2F_{1}\left(1-\beta, 2-\beta; 3-\beta; -\frac{GH^2\beta}{\pi\alpha_+}\right) \right] \nonumber \\
&\, + \frac{3H^2\beta}{16\pi^2\gamma} \left[ \frac{2 \beta HH'}{3\left(2+\beta\right)} \left(\frac{GH^2\beta}{\pi\alpha_-}\right)^\beta
{}_2F_{1}\left(1+\beta, 2+\beta; 3+\beta; -\frac{GH^2\beta}{\pi\alpha_-}\right) \right. \nonumber \\ 
&\, \left. - \frac{2\beta HH'}{3\left(2-\beta\right)}
\left(\frac{GH^2\beta}{\pi\alpha_+}\right)^{-\beta}
{}_2F_{1}\left(1-\beta, 2-\beta; 3-\beta; -\frac{GH^2\beta}{\pi\alpha_+}\right) \right] \nonumber \\
&\, - \frac{H^3 H' \beta}{8\pi^2\gamma} \left[ \frac{1+\beta}{3+\beta} \left(\frac{GH^2\beta}{\pi\alpha_-}\right)^{\beta+1}
{}_2F_{1}\left(2+\beta, 3+\beta; 4+\beta; -\frac{GH^2\beta}{\pi\alpha_-}\right) \right. \nonumber\\ 
&\, \left. + \frac{1-\beta}{3-\beta} \left(\frac{GH^2\beta}{\pi\alpha_+}\right)^{-\beta+1}
{}_2F_{1}\left(2-\beta, 3-\beta; 4-\beta; -\frac{GH^2\beta}{\pi\alpha_+}\right) \right] \, . 
\end{align}
By replacing $H$ by a function $\eta_N(\phi)$ in \eqref{enomgV}, $\omega$ and $V$ are expressed as functions of $\phi$. 
In the obtained model, a solution is given by $\phi=N$ and $H=\eta_N(N)$ as in \eqref{solN}, again. 
Especially when we choose \eqref{HubbleNewCN}, the Hubble rate $H$ in \eqref{HubbleNewC} is reproduced, and by the tuning of the parameters in \eqref{HubbleNewC}, the constraints by ACT are satisfied in the obtained model. 
Just in the same way, one can easily obtain ACT-consistent inflation for any generalised entropy from Section~\ref{Sec6}. 
This is in contrast with pure entropic gravity, where it was observed that it is rather hard to obtain an entropic universe leading to ACT-consistent inflation.

\section{Summury and Discussion\label{Sec7}}

The entropic cosmology consistent with the ACT constraints \eqref{nsACTonly} or \eqref{nsr} \cite{AtacamaCosmologyTelescope:2025nti} by using the generalised entropies has been constructed as in \eqref{Gf3} and \eqref{Gf2C}. 
These entropic cosmology models correspond to the $f(Q)$ gravity \eqref{arbQinf} and \eqref{arbCQ}, which describe cosmology consistent with the ACT results. 
For the formulation of the $f(Q)$ gravity models, we used the reconstruction formulation proposed in \cite{Nojiri:2024zab}. 
One model was reconstructed by using the cosmological time, and another model was reconstructed by the $e$-folding number. 
One-to-one correspondence, up to a constant, between the generalised entropies and the $f(Q)$ gravity, as in \eqref{gSeqFQ2} or \eqref{Gf2}, has been established in \cite{Nojiri:2025fiu}. 
By using the correspondence, we constructed generalised entropies, which give the ACT-consistent inflationary entropic cosmology. 
Within $f(Q)$ gravity coupled with scalar and its equivalent entropic gravity coupled with scalar, we demonstrated how to obtain the arbitrary cosmic evolution tuning corresponding scalar potentials.
It is explicitly demonstrated that the universe based on generalised entropy with several parameters may give ACT-consistent inflation. 
In other words, by tuning scalar potentials, one can obtain realistic ACT inflation for any of the generalised entropies discussed in Section~\ref{Sec6}. 
Note that without scalar it was rather hard to find entropic gravity, which gives ACT inflation. 
The models with the coupling between a scalar field and the entropic cosmology based on the generalised entropies were also constructed. 

As discussed at the end of the last section, a problem could be the reheating or particle production at the end of the inflation. 
Because there is no inflaton, we cannot expect the particle creation by the inflaton decay or its analogues. 
A possible way is the particle creation due to the quantum effects associated with the expansion of the universe. 
The quantity of the created energy density by the quantum effects could be much smaller than that by the inflaton decay, but because the anti-particles are also created, the number of particles remaining in the present universe is very small compared with the number of particles created after the inflation, and there could not be manifest conflicts with observations. 
The number of particles remaining in the present universe must be consistent with baryogenesis.
Eventually, this problem is solved when we add scalar theory to consideration. 

Another way in $f(Q)$ gravity theory side could be rewrite Eq.~\eqref{vQ_FLRW0} in the following form, 
\begin{align}
\label{vQ_FLRW0R1}
\frac{3}{\kappa^2} H^2 =&\, \rho_\mathrm{total} \equiv \rho + \rho_Q \, , \nonumber \\
\rho_Q \equiv &\, - \frac{f(Q)}{2} + Q f'(Q) - \frac{Q}{2\kappa^2}\, , \nonumber \\
-\frac{1}{\kappa^2} \left( 2\dot H + 3 H^2 \right) =&\, p_\mathrm{total} = p + p_Q \, , \nonumber \\
p_Q \equiv&\, \frac{f(Q)}{2} - Q f'(Q) - \frac{\dot Q}{\sqrt{ - 6 Q}} f'(Q) + 4 Q \dot H f''(Q) -\frac{1}{\kappa^2} \left( - \frac{\dot Q}{\sqrt{ - 6 Q}} - \frac{Q}{2} \right) \, .
\end{align}
Then due to Bianchi identity, $\rho_\mathrm{total}$ and $p_\mathrm{total}$ always satisfy the conservation law, 
\begin{align}
\label{ttlconsv}
0 = \dot\rho_\mathrm{total} + 3 H \left( \rho_\mathrm{total} + p_\mathrm{total} \right) \, . 
\end{align}
If the energy density $\rho$ and the pressure $p$ of matter satisfy the conservation law, $\rho_Q$ and $p_Q$ also satisfy the conservation law. 
We may consider, however, the interaction between the matter and $\rho_Q$ or $p_Q$, 
\begin{align}
\label{intconsv}
\dot\rho + 3 H \left( \rho + p \right) = \Gamma (Q)\, , \quad 
\dot\rho_Q + 3 H \left( \rho_Q + p_Q \right) = - \Gamma (Q)\, .
\end{align}
Then, although the total energy and pressure are conserved, there is a transition of the energy between the matter and $\rho_Q$ or $p_Q$. 
As in \eqref{slwrllprmtrs} and \eqref{endN}, we may define the end of the inflation by $\epsilon=1$. 
Let the value of $H$ when $\epsilon=1$ be $H_\mathrm{end}$. 
Then the $\rho_Q$ can be estimated to be 
\begin{align}
\label{rhoQend}
\rho_Q = \left. \left\{ - \frac{f(Q)}{2} + Q f'(Q) - \frac{Q}{2\kappa^2}\right\}\right|_{Q=-6 {H_\mathrm{end}}^2}\, .
\end{align}
The energy may be converted to the particle creation via the interaction in \eqref{intconsv}. 
The reheating temperature could be given by the Hawking temperature $T_\mathrm{H}$ in \eqref{Tslls6} when $\epsilon=1$. 
See \cite{Odintsov:2025bmp} for a similar consideration. 

About the reheating, it could be interesting to reconsider the thermodynamics of the generalised entropy. 
The first law of thermodynamics is also known as $T_\mathrm{H}dS = d\mathcal{Q}$. 
Here $T_\mathrm{H}$ is the temperature in \eqref{Tslls6} and $\mathcal{Q}$ is the heat in \eqref{Tslls2}. 
The heat usually corresponds to the heat flux penetrating the horizon, but it might include the heat of the particle created by the reheating. 
The construction of thermodynamics, including particle production as in the quantum effects, might solve the problem of reheating. 

If the reheating mechanism differs from that of the inflaton decay, the spectrum of the created particles could change, affecting big-bang baryogenesis, dark matter production, primordial gravitational waves, etc. 

As an extension of $f(Q)$ gravity, or $f(Q,C)$ \cite{Gadbail:2023mvu} and equivalently $f(Q,B)$ gravities \cite{Capozziello:2023vne}, we may consider $f(Q, \mathcal{G})$ gravity~\cite{Nojiri:2024hau} whose action is given by, 
\begin{align}
\label{fQG1}
S_{Q\mathcal{G}}= \int d^4 x \sqrt{-g} f\left(Q, \mathcal{G} \right) \, .
\end{align}
Here $\mathcal{G}$ is the Gauss-Bonnet invariant given by, 
\begin{align}
\label{GB} 
\mathcal{G}={\tilde R}^2 -4 {\tilde R}_{\mu\nu} {\tilde R}^{\mu\nu} + {\tilde R}_{\mu\nu\xi\sigma} {\tilde R}^{\mu\nu\xi\sigma}\, . 
\end{align}
We consider the FLRW universe with a flat spatial part \eqref{FLRW}. 
When the scale factor $a(t)$ and therefore the Hubble rate $H\equiv \frac{\dot a}{a}$ is given by a function of the cosmological time $t$, $a=a(t)$ and $H=H(t)$, because $Q=-6H^2$, $Q$ is also given by a function ot $t$, $Q=Q(t)$, which could be solved with respect to $t$, $t=t(Q)$. 
By using this expression, we can express $\mathcal{G}$, $\dot{\mathcal{G}}$, and $\rho$ as the functions of $Q$, $\mathcal{G}=\mathcal{G}(Q)$, $\dot{\mathcal{G}}=\dot{\mathcal{G}}(Q)$, and $\rho=\rho(Q)$. 
For a simpler model, 
\begin{align}
\label{fQG3}
f = f_1(Q) + f_2 \left( \mathcal{G} \right)\, .
\end{align}
The $f_1(Q)$ corresponding to the given scale factor $a(t)$ can be found, 
\begin{align}
\label{fQG5} 
f_1(Q) = \sqrt{-Q} \int^Q \frac{dq}{\sqrt{-q}} \left\{ - f_2\left( \mathcal{G} \left( q \right) \right) 
+ \mathcal{G} \left( q \right) f'_2 \left( \mathcal{G} \left( q \right) \right) 
+ \frac{4}{\sqrt{6}} \dot{\mathcal{G}} \left( q \right) f''_2 \left( \mathcal{G} \left( q \right) \right) \left( - q\right)^\frac{3}{2} 
+ \rho \left( q \right) \right\}\, .
\end{align}
Therefore, the theory is obtained that realises the given FLRW spacetime \eqref{FLRW}. 
By using \eqref{fQG5}, we can construct models of $f(Q, \mathcal{G})$ gravity consistent with the ACT constraints. 
This will be considered elsewhere.

\section*{Acknowledgments}

This research was carried out whilst S.N. was a visiting researcher at the Tsung-Dao Lee Institute, Shanghai Jiao Tong University. 
S.N. is deeply grateful for the tremendous support and warm hospitality received from the Institute, and in particular from Professors Xiao-Gang He and Hong-Jian He. 
S.N. would also like to express his gratitude to Professor Hong-Jian He for his invitation and for the opportunity to engage in discussions with him.
SO acknowledges support from the Arrakihs mission through grant number PID2025-168567OB-C62 of the Spanish Ministerio de Ciencia, Innovaci\'{o}n y Universidades.

\section*{Statements}

No new data were used for this manuscript.

\bibliographystyle{apsrev4-1}
\bibliography{References4}

@article{Nojiri:2006je,
    author = "Nojiri, Shin'ichi and Odintsov, Sergei D. and Sami, M.",
    title = "{Dark energy cosmology from higher-order, string-inspired gravity and its reconstruction}",
    eprint = "hep-th/0605039",
    archivePrefix = "arXiv",
    doi = "10.1103/PhysRevD.74.046004",
    journal = "Phys. Rev. D",
    volume = "74",
    pages = "046004",
    year = "2006"
}

@article{Bekenstein:1973ur,
    author = "Bekenstein, Jacob D.",
    title = "{Black holes and entropy}",
    doi = "10.1103/PhysRevD.7.2333",
    journal = "Phys. Rev. D",
    volume = "7",
    pages = "2333--2346",
    year = "1973"
}

@article{Hawking:1975vcx,
    author = "Hawking, S. W.",
    editor = "Gibbons, G. W. and Hawking, S. W.",
    title = "{Particle Creation by Black Holes}",
    doi = "10.1007/BF02345020",
    journal = "Commun. Math. Phys.",
    volume = "43",
    pages = "199--220",
    year = "1975",
    note = "[Erratum: Commun.Math.Phys. 46, 206 (1976)]"
}

@article{Nojiri:2022aof,
    author = "Nojiri, Shin'ichi and Odintsov, Sergei D. and Faraoni, Valerio",
    title = "{From nonextensive statistics and black hole entropy to the holographic dark universe}",
    eprint = "2201.02424",
    archivePrefix = "arXiv",
    primaryClass = "gr-qc",
    doi = "10.1103/PhysRevD.105.044042",
    journal = "Phys. Rev. D",
    volume = "105",
    number = "4",
    pages = "044042",
    year = "2022"
}

@article{Nojiri:2022dkr,
    author = "Nojiri, Shin'ichi and Odintsov, Sergei D. and Paul, Tanmoy",
    title = "{Early and late universe holographic cosmology from a new generalized entropy}",
    eprint = "2205.08876",
    archivePrefix = "arXiv",
    primaryClass = "gr-qc",
    doi = "10.1016/j.physletb.2022.137189",
    journal = "Phys. Lett. B",
    volume = "831",
    pages = "137189",
    year = "2022"
}

@article{Odintsov:2022qnn,
    author = "Odintsov, Sergei D. and Paul, Tanmoy",
    title = "{A non-singular generalized entropy and its implications on bounce cosmology}",
    eprint = "2212.05531",
    archivePrefix = "arXiv",
    primaryClass = "gr-qc",
    doi = "10.1016/j.dark.2022.101159",
    journal = "Phys. Dark Univ.",
    volume = "39",
    pages = "101159",
    year = "2023"
}

@article{Jacobson:1995ab,
    author = "Jacobson, Ted",
    title = "{Thermodynamics of space-time: The Einstein equation of state}",
    eprint = "gr-qc/9504004",
    archivePrefix = "arXiv",
    reportNumber = "UMDGR-95-114",
    doi = "10.1103/PhysRevLett.75.1260",
    journal = "Phys. Rev. Lett.",
    volume = "75",
    pages = "1260--1263",
    year = "1995"
}

@article{AtacamaCosmologyTelescope:2025nti,
    author = "Calabrese, Erminia and others",
    collaboration = "Atacama Cosmology Telescope",
    title = "{The Atacama Cosmology Telescope: DR6 constraints on extended cosmological models}",
    eprint = "2503.14454",
    archivePrefix = "arXiv",
    primaryClass = "astro-ph.CO",
    reportNumber = "FERMILAB-PUB-25-0157-PPD",
    doi = "10.1088/1475-7516/2025/11/063",
    journal = "JCAP",
    volume = "11",
    pages = "063",
    year = "2025"
}

@article{AtacamaCosmologyTelescope:2025blo,
    author = "Louis, Thibaut and others",
    collaboration = "Atacama Cosmology Telescope",
    title = "{The Atacama Cosmology Telescope: DR6 power spectra, likelihoods and {\ensuremath{\Lambda}}CDM parameters}",
    eprint = "2503.14452",
    archivePrefix = "arXiv",
    primaryClass = "astro-ph.CO",
    reportNumber = "FERMILAB-PUB-25-0071-PPD",
    doi = "10.1088/1475-7516/2025/11/062",
    journal = "JCAP",
    volume = "11",
    pages = "062",
    year = "2025"
}

@article{Izumi:2012qj,
    author = "Izumi, Keisuke and Ong, Yen Chin",
    title = "{Cosmological Perturbation in f(T) Gravity Revisited}",
    eprint = "1212.5774",
    archivePrefix = "arXiv",
    primaryClass = "gr-qc",
    doi = "10.1088/1475-7516/2013/06/029",
    journal = "JCAP",
    volume = "06",
    pages = "029",
    year = "2013"
}

@article{BeltranJimenez:2019tme,
    author = "Beltr{\'a}n Jim{\'e}nez, Jose and Heisenberg, Lavinia and Koivisto, Tomi Sebastian and Pekar, Simon",
    title = "{Cosmology in $f(Q)$ geometry}",
    eprint = "1906.10027",
    archivePrefix = "arXiv",
    primaryClass = "gr-qc",
    doi = "10.1103/PhysRevD.101.103507",
    journal = "Phys. Rev. D",
    volume = "101",
    number = "10",
    pages = "103507",
    year = "2020"
}

@article{Nojiri:2024zab,
    author = "Nojiri, Shin'ichi and Odintsov, S. D.",
    title = "{Well-defined f(Q) gravity, reconstruction of FLRW spacetime and unification of inflation with dark energy epoch}",
    eprint = "2404.18427",
    archivePrefix = "arXiv",
    primaryClass = "gr-qc",
    reportNumber = "KEK-Cosmo-0344, KEK-TH-2624",
    doi = "10.1016/j.dark.2024.101538",
    journal = "Phys. Dark Univ.",
    volume = "45",
    pages = "101538",
    year = "2024"
}

@article{Heisenberg:2023wgk,
    author = "Heisenberg, Lavinia and Hohmann, Manuel and Kuhn, Simon",
    title = "{Cosmological teleparallel perturbations}",
    eprint = "2311.05495",
    archivePrefix = "arXiv",
    primaryClass = "gr-qc",
    doi = "10.1088/1475-7516/2024/03/063",
    journal = "JCAP",
    volume = "03",
    pages = "063",
    year = "2024"
}

@article{Gomes:2023tur,
    author = "Gomes, D{\'e}bora Aguiar and Beltr{\'a}n Jim{\'e}nez, Jose and Cano, Alejandro Jim{\'e}nez and Koivisto, Tomi S.",
    title = "{Pathological Character of Modifications to Coincident General Relativity: Cosmological Strong Coupling and Ghosts in f(Q) Theories}",
    eprint = "2311.04201",
    archivePrefix = "arXiv",
    primaryClass = "gr-qc",
    doi = "10.1103/PhysRevLett.132.141401",
    journal = "Phys. Rev. Lett.",
    volume = "132",
    number = "14",
    pages = "141401",
    year = "2024"
}

@article{Hu:2023gui,
    author = "Hu, Kun and Yamakoshi, Makishi and Katsuragawa, Taishi and Nojiri, Shin'ichi and Qiu, Taotao",
    title = "{Nonpropagating ghost in covariant f(Q) gravity}",
    eprint = "2310.15507",
    archivePrefix = "arXiv",
    primaryClass = "gr-qc",
    doi = "10.1103/PhysRevD.108.124030",
    journal = "Phys. Rev. D",
    volume = "108",
    number = "12",
    pages = "124030",
    year = "2023"
}

@article{Nojiri:2025fiu,
    author = "Nojiri, Shin'ichi and Odintsov, Sergei D.",
    title = "{The correspondence of generalised entropic cosmology theory with F(T) and F(Q) modified gravity and gravitational waves}",
    eprint = "2502.15272",
    archivePrefix = "arXiv",
    primaryClass = "gr-qc",
    reportNumber = "KEK-TH-2690, KEK-Cosmo-0373",
    doi = "10.1016/j.dark.2025.101899",
    journal = "Phys. Dark Univ.",
    volume = "48",
    pages = "101899",
    year = "2025"
}

@article{Ong:2013qja,
    author = "Ong, Yen Chin and Izumi, Keisuke and Nester, James M. and Chen, Pisin",
    title = "{Problems with Propagation and Time Evolution in f(T) Gravity}",
    eprint = "1303.0993",
    archivePrefix = "arXiv",
    primaryClass = "gr-qc",
    doi = "10.1103/PhysRevD.88.024019",
    journal = "Phys. Rev. D",
    volume = "88",
    pages = "024019",
    year = "2013"
}

@article{Tsallis:1987eu,
    author = "Tsallis, Constantino",
    title = "{Possible Generalization of Boltzmann-Gibbs Statistics}",
    reportNumber = "CBPF-NF-062-87",
    doi = "10.1007/BF01016429",
    journal = "J. Statist. Phys.",
    volume = "52",
    pages = "479--487",
    year = "1988"
}

@article{Kaniadakis:2005zk,
    author = "Kaniadakis, G.",
    title = "{Statistical mechanics in the context of special relativity. II.}",
    eprint = "cond-mat/0507311",
    archivePrefix = "arXiv",
    doi = "10.1103/PhysRevE.72.036108",
    journal = "Phys. Rev. E",
    volume = "72",
    pages = "036108",
    year = "2005"
}

@article{Drepanou:2021jiv,
    author = "Drepanou, Niki and Lymperis, Andreas and Saridakis, Emmanuel N. and Yesmakhanova, Kuralay",
    title = "{Kaniadakis holographic dark energy and cosmology}",
    eprint = "2109.09181",
    archivePrefix = "arXiv",
    primaryClass = "gr-qc",
    doi = "10.1140/epjc/s10052-022-10415-9",
    journal = "Eur. Phys. J. C",
    volume = "82",
    number = "5",
    pages = "449",
    year = "2022"
}

@article{Majhi:2017zao,
    author = "Majhi, Abhishek",
    title = "{Non-extensive Statistical Mechanics and Black Hole Entropy From Quantum Geometry}",
    eprint = "1703.09355",
    archivePrefix = "arXiv",
    primaryClass = "gr-qc",
    doi = "10.1016/j.physletb.2017.10.043",
    journal = "Phys. Lett. B",
    volume = "775",
    pages = "32--36",
    year = "2017"
}

@article{Liu:2021dvj,
    author = "Liu, Yang",
    title = "{Non-extensive statistical mechanics and the thermodynamic stability of FRW universe}",
    eprint = "2112.15077",
    archivePrefix = "arXiv",
    primaryClass = "gr-qc",
    doi = "10.1209/0295-5075/ac3f52",
    journal = "EPL",
    volume = "138",
    number = "3",
    pages = "39001",
    year = "2022"
}

@inproceedings{Renyi,
    author = "R{\'{e}}nyi, A",
    title = "{On measures of information and entropy}",
    booktitle = "{Proceedings of the Fourth Berkeley Symposium on Mathematics, Statistics 
and Probability}",
    pages = "547--56",
    year = "1960",
    publisher = "University of California Press"
}

@article{Barrow:2020tzx,
    author = "Barrow, John D.",
    title = "{The Area of a Rough Black Hole}",
    eprint = "2004.09444",
    archivePrefix = "arXiv",
    primaryClass = "gr-qc",
    doi = "10.1016/j.physletb.2020.135643",
    journal = "Phys. Lett. B",
    volume = "808",
    pages = "135643",
    year = "2020"
}

@article{SayahianJahromi:2018irq,
    author = "Sayahian Jahromi, A. and Moosavi, S. A. and Moradpour, H. and Morais Gra{\c{c}}a, J. P. and Lobo, I. P. and Salako, I. G. and Jawad, A.",
    title = "{Generalized entropy formalism and a new holographic dark energy model}",
    eprint = "1802.07722",
    archivePrefix = "arXiv",
    primaryClass = "gr-qc",
    doi = "10.1016/j.physletb.2018.02.052",
    journal = "Phys. Lett. B",
    volume = "780",
    pages = "21--24",
    year = "2018"
}

@article{SharmaMittal1,
    author = "Sharma, B. D. and Mittal, D. P.",
    title = "{New Non-additive Measures of Entropy for Discrete Probability Distributions}",
    journal = "Journal of Mathematical Sciences",
    volume = "10",
    pages = "28--40, MR 0539493",
    year = "1975"
}

@article{SharmaMittal2,
    author = "Sharma, B. D. and Mittal, D. P.",
    title = "{New Non-additive Measures of Entropy for Discrete Probability Distributions}",
    journal = "Journal of Combinatorics, Information \& System Sciences",
    volume = "2",
    pages = "122",
    year = "1977"
}

@article{Cai:2005ra,
    author = "Cai, Rong-Gen and Kim, Sang Pyo",
    title = "{First law of thermodynamics and Friedmann equations of Friedmann-Robertson-Walker universe}",
    eprint = "hep-th/0501055",
    archivePrefix = "arXiv",
    doi = "10.1088/1126-6708/2005/02/050",
    journal = "JHEP",
    volume = "02",
    pages = "050",
    year = "2005"
}

@article{Akbar:2006er,
    author = "Akbar, M. and Cai, Rong-Gen",
    title = "{Friedmann equations of FRW universe in scalar-tensor gravity, f(R) gravity and first law of thermodynamics}",
    eprint = "hep-th/0602156",
    archivePrefix = "arXiv",
    doi = "10.1016/j.physletb.2006.02.035",
    journal = "Phys. Lett. B",
    volume = "635",
    pages = "7--10",
    year = "2006"
}

@article{Nojiri:2026hij,
    author = "Nojiri, Shin'ichi and Odintsov, Sergei and Oikonomou, V. K.",
    title = "{Ghost-free non-local F(R) gravity compatible with ACT}",
    eprint = "2601.07879",
    archivePrefix = "arXiv",
    primaryClass = "gr-qc",
    reportNumber = "KEK-TH-2800, KEK-Cosmo-0407",
    doi = "10.1016/j.physletb.2026.140290",
    journal = "Phys. Lett. B",
    volume = "874",
    pages = "140290",
    year = "2026"
}

@article{Odintsov:2025eiv,
    author = "Odintsov, S. D. and Oikonomou, V. K.",
    title = "{Power-law F(R) gravity as deformations to Starobinsky inflation in view of ACT}",
    eprint = "2509.06251",
    archivePrefix = "arXiv",
    primaryClass = "gr-qc",
    doi = "10.1016/j.physletb.2025.139907",
    journal = "Phys. Lett. B",
    volume = "870",
    pages = "139907",
    year = "2025"
}

@article{Yang:2026rzn,
    author = "Yang, Rui and Tao, Jun and Wang, Peng and Zhu, Mian",
    title = "{Revisiting the Lyth bound constraints on inflation from ACT DR6 results}",
    eprint = "2606.16711",
    archivePrefix = "arXiv",
    primaryClass = "astro-ph.CO",
    month = "6",
    year = "2026"
}

@article{Gonuguntla:2026rkw,
    author = "Gonuguntla, Haneesh and Modak, Tanmoy and Samanta, Arnab",
    title = "{Nonminimal couplings and preheating effects in $R^2$-Higgs inflation after ACT and SPT}",
    eprint = "2606.11929",
    archivePrefix = "arXiv",
    primaryClass = "astro-ph.CO",
    month = "6",
    year = "2026"
}

@article{CastroJunior:2026gnm,
    author = "Castro Junior, A. Oliveira and Monerat, G. A. and Oliveira-Neto, G. and Corr{\^e}a Silva, E. V.",
    title = "{Einstein{\textendash}Aether primordial universe with radiation and dark energy}",
    doi = "10.1140/epjc/s10052-026-15788-9",
    journal = "Eur. Phys. J. C",
    volume = "86",
    number = "5",
    pages = "565",
    year = "2026"
}

@article{Odintsov:2026dss,
    author = "Odintsov, S. D. and Oikonomou, V. K.",
    title = "{Positive Running of the Spectral Index for Scalar Theory and Modified Gravity}",
    eprint = "2605.17813",
    archivePrefix = "arXiv",
    primaryClass = "gr-qc",
    month = "5",
    year = "2026"
}

@article{Yang:2026flt,
    author = "Yang, Wei and Wu, Chen-Hao and Hu, Ya-Peng",
    title = "{Constraints on non-canonical chaotic inflation from ACT DR6 and BICEP/Keck data}",
    eprint = "2605.16772",
    archivePrefix = "arXiv",
    primaryClass = "gr-qc",
    doi = "10.1016/j.physletb.2026.140532",
    journal = "Phys. Lett. B",
    volume = "878",
    pages = "140532",
    year = "2026"
}

@article{Zambrano:2026fau,
    author = "Zambrano, Jordan and Rodr{\'\i}guez, {\'A}ngel and Alvear, Melisa and Rojas, Clara",
    title = "{Numerical Study of Some Generalizations of the Starobinsky Inflationary Model}",
    eprint = "2605.10728",
    archivePrefix = "arXiv",
    primaryClass = "astro-ph.CO",
    month = "5",
    year = "2026"
}

@article{Latosh:2026ckf,
    author = "Latosh, Boris",
    title = "{Robustness of Starobinsky inflation in a minimal two-field scalar-tensor completion}",
    eprint = "2604.15931",
    archivePrefix = "arXiv",
    primaryClass = "gr-qc",
    month = "4",
    year = "2026"
}

@article{Yogesh:2026esn,
    author = "Yogesh and Bhat, Imtiyaz Ahmad and Gangopadhyay, Mayukh R. and Sami, M.",
    title = "{Constraining Quintessential Inflation with ACT: A Gauss-Bonnet Gateway}",
    eprint = "2604.14659",
    archivePrefix = "arXiv",
    primaryClass = "astro-ph.CO",
    month = "4",
    year = "2026"
}

@article{Ahmed:2026agd,
    author = "Ahmed, Waqas and Allehabi, Saleh O. and Rehman, Mansoor Ur",
    title = "{Radiatively Corrected Hybrid Inflation: Parameter Scans and Machine Learning with ACT and Future CMB Experiments}",
    eprint = "2604.11068",
    archivePrefix = "arXiv",
    primaryClass = "hep-ph",
    month = "4",
    year = "2026"
}

@article{Whittingham:2026cbo,
    author = "Whittingham, Ian B.",
    title = "{Gravitational Baryogenesis in $f(R)$ Cosmologies}",
    eprint = "2602.19075",
    archivePrefix = "arXiv",
    primaryClass = "astro-ph.CO",
    month = "2",
    year = "2026"
}

@article{Yuennan:2026fcn,
    author = "Yuennan, Jureeporn and Atamurotov, Farruh and Capozziello, Salvatore and Channuie, Phongpichit",
    title = "{Constraining $\beta $-exponential inflation with the latest ACT observations}",
    eprint = "2602.17380",
    archivePrefix = "arXiv",
    primaryClass = "gr-qc",
    doi = "10.1140/epjc/s10052-026-15461-1",
    journal = "Eur. Phys. J. C",
    volume = "86",
    number = "3",
    pages = "237",
    year = "2026"
}

@article{Ahmed:2026msg,
    author = "Ahmed, Waqas and Ahmad, Waqar and Illahi, Ahsan and Junaid, M.",
    title = "{Warm Hybrid Axion Inflation in $α$-Attractor Models Constrained by ACT and Future Plan experiments}",
    eprint = "2601.10145",
    archivePrefix = "arXiv",
    primaryClass = "hep-ph",
    month = "1",
    year = "2026"
}

@article{Peng:2026ofs,
    author = "Peng, Ze-Yu and Yuan, Hao-Shi and Lai, Qi and Lan, Qing-Yu and Wang, Zhan-He and Jiang, Jun-Qian and Ye, Gen and Zhang, Jun and Piao, Yun-Song",
    title = "{DeepInflation: An AI Agent for Research and Model Discovery of Inflation}",
    eprint = "2601.14288",
    archivePrefix = "arXiv",
    primaryClass = "astro-ph.CO",
    doi = "10.1088/1674-4527/ae6ad8",
    journal = "Res. Astron. Astrophys.",
    volume = "26",
    number = "11",
    pages = "115020",
    year = "2026"
}

@article{Chattopadhyay:2026yam,
    author = "Chattopadhyay, Gahan and Sengupta, Soumitra",
    title = "{Emergence of Einstein's gravity from higher curvature f(R) theories through cosmological evolution}",
    eprint = "2601.01395",
    archivePrefix = "arXiv",
    primaryClass = "gr-qc",
    month = "1",
    year = "2026"
}

@article{Modak:2025grj,
    author = "Modak, Tanmoy",
    title = "{Echoes of R3 modification and Goldstone preheating in the CMB-BAO landscape}",
    eprint = "2512.20730",
    archivePrefix = "arXiv",
    primaryClass = "hep-ph",
    doi = "10.1016/j.physletb.2026.140384",
    journal = "Phys. Lett. B",
    volume = "876",
    pages = "140384",
    year = "2026"
}

@article{Thakur:2025uua,
    author = "Thakur, Rahul and Ajith, Abhijith and Panda, Sukanta and Vidyarthi, Archit",
    title = "{Phantom Menace in general Palatini $f(R,ϕ)$ theories}",
    eprint = "2512.16256",
    archivePrefix = "arXiv",
    primaryClass = "gr-qc",
    month = "12",
    year = "2025"
}

@article{Wang:2025cpp,
    author = "Wang, Qing-Yang",
    title = "{Inflation in light of ACT/SPT: A new perspective from Weyl gravity}",
    eprint = "2512.10862",
    archivePrefix = "arXiv",
    primaryClass = "astro-ph.CO",
    doi = "10.1103/842x-zhtl",
    journal = "Phys. Rev. D",
    volume = "113",
    number = "12",
    pages = "L121302",
    year = "2026"
}

@article{McDonough:2025lzo,
    author = "McDonough, Evan and Ferreira, Elisa G. M.",
    title = "{The spectrum of $n_s$ constraints from DESI and CMB data}",
    eprint = "2512.05108",
    archivePrefix = "arXiv",
    primaryClass = "astro-ph.CO",
    month = "12",
    year = "2025"
}

@article{Keskin:2025zqq,
    author = "Keskin, A. I.",
    title = "{Rainbow f(R) Inflation: de Sitter-Quintessence Phases}",
    doi = "10.1016/j.physletb.2025.140069",
    journal = "Phys. Lett. B",
    volume = "872",
    pages = "140069",
    year = "2026"
}

@article{Chakraborty:2025wqn,
    author = "Chakraborty, Dibya and Hai, Mishaal and Jahan, Sayeda Tashnuba and Kamal, Ahmed Rakin and Shuvo, Md Shaikot Jahan",
    title = "{Effect of moduli redefinitions on fibre inflation}",
    eprint = "2511.19610",
    archivePrefix = "arXiv",
    primaryClass = "hep-th",
    doi = "10.1088/1475-7516/2026/06/025",
    journal = "JCAP",
    volume = "06",
    pages = "025",
    year = "2026"
}

@article{Yuennan:2025mlg,
    author = "Yuennan, Jureeporn and Eadkhong, Thammarong and Atamurotov, Farruh and Channuie, Phongpichit",
    title = "{Constraining non-minimally coupled squared-quartic hilltop inflation in light of ACT observations}",
    eprint = "2511.17216",
    archivePrefix = "arXiv",
    primaryClass = "astro-ph.CO",
    doi = "10.1016/j.dark.2026.102282",
    journal = "Phys. Dark Univ.",
    volume = "52",
    pages = "102282",
    year = "2026"
}

@article{Bezerra-Sobrinho:2025gfg,
    author = "Bezerra-Sobrinho, J. and Medeiros, L. G.",
    title = "{Starobinsky inflation and the latest CMB data: a subtle tension?}",
    eprint = "2511.06640",
    archivePrefix = "arXiv",
    primaryClass = "astro-ph.CO",
    doi = "10.1140/epjc/s10052-026-15686-0",
    journal = "Eur. Phys. J. C",
    volume = "86",
    number = "4",
    pages = "416",
    year = "2026"
}

@article{Fu:2025ciy,
    author = "Fu, Chengjie and Lu, Di and Wang, Shao-Jiang",
    title = "{Harrison-Zeldovich attractor: From Planck to ACT results}",
    eprint = "2510.24682",
    archivePrefix = "arXiv",
    primaryClass = "astro-ph.CO",
    doi = "10.1103/cm1c-kfpn",
    journal = "Phys. Rev. D",
    volume = "113",
    number = "8",
    pages = "L081304",
    year = "2026"
}

@article{Afshar:2025ndm,
    author = "Afshar, Mohammad Ali S. and Noori Gashti, Saeed and Alipour, Mohammad Reza and Pourhassan, Behnam and Sakalli, Izzet and Sadeghi, Jafar",
    title = "{Swampland Conjectures through ACT Observations: Observational Signatures of Radiative-Corrected Inflation}",
    eprint = "2510.20876",
    archivePrefix = "arXiv",
    primaryClass = "astro-ph.CO",
    month = "10",
    year = "2025"
}

@article{Qiu:2025iqm,
    author = "Qiu, Zhichong and Pang, Yehuang and Huang, Qingguo",
    title = "{The implications of inflation for the last ACT}",
    eprint = "2510.18320",
    archivePrefix = "arXiv",
    primaryClass = "astro-ph.CO",
    doi = "10.1007/s11433-025-2934-8",
    journal = "Sci. China Phys. Mech. Astron.",
    volume = "69",
    number = "6",
    pages = "260413",
    year = "2026"
}

@article{Yuennan:2025tyx,
    author = "Yuennan, Jureeporn and Atamurotov, Farruh and Channuie, Phongpichit",
    title = "{ACT constraints on marginally deformed starobinsky inflation}",
    eprint = "2509.23329",
    archivePrefix = "arXiv",
    primaryClass = "gr-qc",
    doi = "10.1016/j.physletb.2025.140065",
    journal = "Phys. Lett. B",
    volume = "872",
    pages = "140065",
    year = "2026"
}

@article{Pozdeeva:2025wsl,
    author = "Pozdeeva, E. O. and Vernov, S. Yu.",
    title = "{Primordial Black Holes Formation in Inflationary $\boldsymbol{F(R)}$ Models with Scalar Fields}",
    eprint = "2509.21220",
    archivePrefix = "arXiv",
    primaryClass = "gr-qc",
    doi = "10.3103/S0027134925702753",
    journal = "Moscow Univ. Phys. Bull.",
    volume = "80",
    number = "Suppl 2",
    pages = "S903--S912",
    year = "2025"
}

@article{Hell:2025lbl,
    author = "Hell, Anamaria and Lust, Dieter",
    title = "{Aspects of non-minimally coupled curvature with power laws}",
    eprint = "2509.20217",
    archivePrefix = "arXiv",
    primaryClass = "hep-th",
    reportNumber = "IPMU25-0046, LMU-ASC 22/25, MPP-2025-187",
    doi = "10.1007/JHEP12(2025)091",
    journal = "JHEP",
    volume = "12",
    pages = "091",
    year = "2025"
}

@article{Zhu:2025twm,
    author = "Zhu, Yigan and Gao, Qing and Gong, Yungui and Yi, Zhu",
    title = "{Inflationary models with Gauss{\textendash}Bonnet coupling in light of ACT observations}",
    eprint = "2508.09707",
    archivePrefix = "arXiv",
    primaryClass = "astro-ph.CO",
    doi = "10.1140/epjc/s10052-025-14969-2",
    journal = "Eur. Phys. J. C",
    volume = "85",
    number = "10",
    pages = "1227",
    year = "2025"
}

@article{Ahmed:2025sfm,
    author = "Ahmed, Waqas and Allehabi, Saleh O. and Rehman, Mansoor Ur",
    title = "{Revisiting polynomial hybrid inflation: Planck and ACT compatibility via radiative corrections}",
    eprint = "2508.01998",
    archivePrefix = "arXiv",
    primaryClass = "hep-ph",
    doi = "10.1103/jxg5-khj2",
    journal = "Phys. Rev. D",
    volume = "113",
    number = "4",
    pages = "043532",
    year = "2026"
}

@article{Yi:2025dms,
    author = "Yi, Zhu and Wang, Xingzhi and Gao, Qing and Gong, Yungui",
    title = "{Approximate reconstruction of inflationary potential with ACT observations}",
    eprint = "2505.10268",
    archivePrefix = "arXiv",
    primaryClass = "astro-ph.CO",
    doi = "10.1016/j.physletb.2025.140002",
    journal = "Phys. Lett. B",
    volume = "871",
    pages = "140002",
    year = "2025"
}

@article{Nojiri:2024zdu,
    author = "Nojiri, Shin'ichi and Odintsov, Sergei D. and Paul, Tanmoy",
    title = "{Different Aspects of Entropic Cosmology}",
    eprint = "2409.01090",
    archivePrefix = "arXiv",
    primaryClass = "gr-qc",
    doi = "10.3390/universe10090352",
    journal = "Universe",
    volume = "10",
    number = "9",
    pages = "352",
    year = "2024"
}

@article{Planck:2013jfk,
    author = "Ade, P. A. R. and others",
    collaboration = "Planck",
    title = "{Planck 2013 results. XXII. Constraints on inflation}",
    eprint = "1303.5082",
    archivePrefix = "arXiv",
    primaryClass = "astro-ph.CO",
    reportNumber = "CERN-PH-TH-2013-135",
    doi = "10.1051/0004-6361/201321569",
    journal = "Astron. Astrophys.",
    volume = "571",
    pages = "A22",
    year = "2014"
}

@article{Planck:2015sxf,
    author = "Ade, P. A. R. and others",
    collaboration = "Planck",
    title = "{Planck 2015 results. XX. Constraints on inflation}",
    eprint = "1502.02114",
    archivePrefix = "arXiv",
    primaryClass = "astro-ph.CO",
    doi = "10.1051/0004-6361/201525898",
    journal = "Astron. Astrophys.",
    volume = "594",
    pages = "A20",
    year = "2016"
}

@article{Planck:2018jri,
    author = "Akrami, Y. and others",
    collaboration = "Planck",
    title = "{Planck 2018 results. X. Constraints on inflation}",
    eprint = "1807.06211",
    archivePrefix = "arXiv",
    primaryClass = "astro-ph.CO",
    doi = "10.1051/0004-6361/201833887",
    journal = "Astron. Astrophys.",
    volume = "641",
    pages = "A10",
    year = "2020"
}

@article{Nester:1998mp,
    author = "Nester, James M. and Yo, Hwei-Jang",
    title = "{Symmetric teleparallel general relativity}",
    eprint = "gr-qc/9809049",
    archivePrefix = "arXiv",
    reportNumber = "NCU-CCS-980904",
    journal = "Chin. J. Phys.",
    volume = "37",
    pages = "113",
    year = "1999"
}

@article{BeltranJimenez:2018vdo,
    author = "Beltr{\'a}n Jim{\'e}nez, Jose and Heisenberg, Lavinia and Koivisto, Tomi S.",
    title = "{Teleparallel Palatini theories}",
    eprint = "1803.10185",
    archivePrefix = "arXiv",
    primaryClass = "gr-qc",
    reportNumber = "NORDITA-2018-023, IFT-UAM/CSIC-18-035, IFT-UAM-CSIC-18-035",
    doi = "10.1088/1475-7516/2018/08/039",
    journal = "JCAP",
    volume = "08",
    pages = "039",
    year = "2018"
}

@article{Runkla:2018xrv,
    author = {R{\"u}nkla, Mihkel and Vilson, Ott},
    title = "{Family of scalar-nonmetricity theories of gravity}",
    eprint = "1805.12197",
    archivePrefix = "arXiv",
    primaryClass = "gr-qc",
    doi = "10.1103/PhysRevD.98.084034",
    journal = "Phys. Rev. D",
    volume = "98",
    number = "8",
    pages = "084034",
    year = "2018"
}

@article{Capozziello:2022tvv,
    author = "Capozziello, Salvatore and Shokri, Mehdi",
    title = "{Slow-roll inflation in f(Q) non-metric gravity}",
    eprint = "2209.06670",
    archivePrefix = "arXiv",
    primaryClass = "gr-qc",
    doi = "10.1016/j.dark.2022.101113",
    journal = "Phys. Dark Univ.",
    volume = "37",
    pages = "101113",
    year = "2022"
}

@article{Blixt:2023kyr,
    author = "Blixt, Daniel and Golovnev, Alexey and Guzman, Maria-Jose and Maksyutov, Ramazan",
    title = "{Geometry and covariance of symmetric teleparallel theories of gravity}",
    eprint = "2306.09289",
    archivePrefix = "arXiv",
    primaryClass = "gr-qc",
    doi = "10.1103/PhysRevD.109.044061",
    journal = "Phys. Rev. D",
    volume = "109",
    number = "4",
    pages = "044061",
    year = "2024"
}

@article{Adak:2018vzk,
    author = "Adak, Muzaffer",
    title = "{Gauge Approach to The Symmetric Teleparallel Gravity}",
    eprint = "1809.01385",
    archivePrefix = "arXiv",
    primaryClass = "gr-qc",
    doi = "10.1142/S0219887818501980",
    journal = "Int. J. Geom. Meth. Mod. Phys.",
    volume = "15",
    number = "12",
    pages = "1850198",
    year = "2018"
}

@article{Tomonari:2023wcs,
    author = "Tomonari, Kyosuke and Bahamonde, Sebastian",
    title = "{Dirac{\textendash}Bergmann analysis and degrees of freedom of coincident f(Q)-gravity}",
    eprint = "2308.06469",
    archivePrefix = "arXiv",
    primaryClass = "gr-qc",
    doi = "10.1140/epjc/s10052-024-12677-x",
    journal = "Eur. Phys. J. C",
    volume = "84",
    number = "4",
    pages = "349",
    year = "2024",
    note = "[Erratum: Eur.Phys.J.C 84, 508 (2024)]"
}

@article{Heisenberg:2023lru,
    author = "Heisenberg, Lavinia",
    title = "{Review on f(Q) gravity}",
    eprint = "2309.15958",
    archivePrefix = "arXiv",
    primaryClass = "gr-qc",
    doi = "10.1016/j.physrep.2024.02.001",
    journal = "Phys. Rept.",
    volume = "1066",
    pages = "1--78",
    year = "2024"
}

@article{DAmbrosio:2023asf,
    author = "D'Ambrosio, Fabio and Heisenberg, Lavinia and Zentarra, Stefan",
    title = "{Hamiltonian Analysis of f(Q)$f(\mathbb {Q})$ Gravity and the Failure of the Dirac{\textendash}Bergmann Algorithm for Teleparallel Theories of Gravity}",
    eprint = "2308.02250",
    archivePrefix = "arXiv",
    primaryClass = "gr-qc",
    doi = "10.1002/prop.202300185",
    journal = "Fortsch. Phys.",
    volume = "71",
    number = "12",
    pages = "2300185",
    year = "2023"
}

@article{Paliathanasis:2023pqp,
    author = "Paliathanasis, A. and Dimakis, N. and Christodoulakis, T.",
    title = "{Minisuperspace description of f(Q)-cosmology}",
    eprint = "2308.15207",
    archivePrefix = "arXiv",
    primaryClass = "gr-qc",
    doi = "10.1016/j.dark.2023.101410",
    journal = "Phys. Dark Univ.",
    volume = "43",
    pages = "101410",
    year = "2024"
}

@article{Dimakis:2021gby,
    author = "Dimakis, N. and Paliathanasis, A. and Christodoulakis, T.",
    title = "{Quantum cosmology in f(Q) theory}",
    eprint = "2108.01970",
    archivePrefix = "arXiv",
    primaryClass = "gr-qc",
    doi = "10.1088/1361-6382/ac2b09",
    journal = "Class. Quant. Grav.",
    volume = "38",
    number = "22",
    pages = "225003",
    year = "2021"
}

@article{Bamba:2013ooa,
    author = "Bamba, Kazuharu and Capozziello, Salvatore and De Laurentis, Mariafelicia and Nojiri, Shin'ichi and S{\'a}ez-G{\'o}mez, Diego",
    title = "{No further gravitational wave modes in $F(T)$ gravity}",
    eprint = "1309.2698",
    archivePrefix = "arXiv",
    primaryClass = "gr-qc",
    doi = "10.1016/j.physletb.2013.10.022",
    journal = "Phys. Lett. B",
    volume = "727",
    pages = "194--198",
    year = "2013"
}

@article{Capozziello:2024vix,
    author = "Capozziello, Salvatore and Capriolo, Maurizio and Nojiri, Shin'ichi",
    title = "{Gravitational waves in f(Q) non-metric gravity via geodesic deviation}",
    eprint = "2401.06424",
    archivePrefix = "arXiv",
    primaryClass = "gr-qc",
    doi = "10.1016/j.physletb.2024.138510",
    journal = "Phys. Lett. B",
    volume = "850",
    pages = "138510",
    year = "2024"
}

@article{Nojiri:2005pu,
    author = "Nojiri, Shin'ichi and Odintsov, Sergei D.",
    title = "{Unifying phantom inflation with late-time acceleration: Scalar phantom-non-phantom transition model and generalized holographic dark energy}",
    eprint = "hep-th/0506212",
    archivePrefix = "arXiv",
    doi = "10.1007/s10714-006-0301-6",
    journal = "Gen. Rel. Grav.",
    volume = "38",
    pages = "1285--1304",
    year = "2006"
}

@article{Capozziello:2005tf,
    author = "Capozziello, S. and Nojiri, S. and Odintsov, S. D.",
    title = "{Unified phantom cosmology: Inflation, dark energy and dark matter under the same standard}",
    eprint = "hep-th/0507182",
    archivePrefix = "arXiv",
    doi = "10.1016/j.physletb.2005.11.012",
    journal = "Phys. Lett. B",
    volume = "632",
    pages = "597--604",
    year = "2006"
}

@article{Nojiri:2005jg,
    author = "Nojiri, Shin'ichi and Odintsov, Sergei D.",
    title = "{Modified Gauss-Bonnet theory as gravitational alternative for dark energy}",
    eprint = "hep-th/0508049",
    archivePrefix = "arXiv",
    doi = "10.1016/j.physletb.2005.10.010",
    journal = "Phys. Lett. B",
    volume = "631",
    pages = "1--6",
    year = "2005"
}

@article{Cognola:2006eg,
    author = "Cognola, Guido and Elizalde, Emilio and Nojiri, Shin'ichi and Odintsov, Sergei D. and Zerbini, Sergio",
    title = "{Dark energy in modified Gauss-Bonnet gravity: Late-time acceleration and the hierarchy problem}",
    eprint = "hep-th/0601008",
    archivePrefix = "arXiv",
    doi = "10.1103/PhysRevD.73.084007",
    journal = "Phys. Rev. D",
    volume = "73",
    pages = "084007",
    year = "2006"
}

@article{Nojiri:2019dwl,
    author = "Nojiri, Shin'ichi and Odintsov, S. D. and Oikonomou, V. K. and Chatzarakis, N. and Paul, Tanmoy",
    title = "{Viable inflationary models in a ghost-free Gauss{\textendash}Bonnet theory of gravity}",
    eprint = "1907.00403",
    archivePrefix = "arXiv",
    primaryClass = "gr-qc",
    doi = "10.1140/epjc/s10052-019-7080-1",
    journal = "Eur. Phys. J. C",
    volume = "79",
    number = "7",
    pages = "565",
    year = "2019"
}

@article{Nojiri:2006gh,
    author = "Nojiri, Shin'ichi and Odintsov, Sergei D.",
    title = "{Modified f(R) gravity consistent with realistic cosmology: From matter dominated epoch to dark energy universe}",
    eprint = "hep-th/0608008",
    archivePrefix = "arXiv",
    doi = "10.1103/PhysRevD.74.086005",
    journal = "Phys. Rev. D",
    volume = "74",
    pages = "086005",
    year = "2006"
}

@article{Nojiri:2009kx,
    author = "Nojiri, Shin'ichi and Odintsov, Sergei D. and Saez-Gomez, Diego",
    title = "{Cosmological reconstruction of realistic modified F(R) gravities}",
    eprint = "0908.1269",
    archivePrefix = "arXiv",
    primaryClass = "hep-th",
    doi = "10.1016/j.physletb.2009.09.045",
    journal = "Phys. Lett. B",
    volume = "681",
    pages = "74--80",
    year = "2009"
}

@article{Capozziello:2022wgl,
    author = "Capozziello, Salvatore and D'Agostino, Rocco",
    title = "{Model-independent reconstruction of f(Q) non-metric gravity}",
    eprint = "2204.01015",
    archivePrefix = "arXiv",
    primaryClass = "gr-qc",
    doi = "10.1016/j.physletb.2022.137229",
    journal = "Phys. Lett. B",
    volume = "832",
    pages = "137229",
    year = "2022"
}

@article{Gadbail:2023klq,
    author = "Gadbail, Gaurav N. and Arora, Simran and Sahoo, P. K.",
    title = "{Reconstruction of f(Q,T) Lagrangian for various cosmological scenario}",
    eprint = "2301.08876",
    archivePrefix = "arXiv",
    primaryClass = "gr-qc",
    doi = "10.1016/j.physletb.2023.137710",
    journal = "Phys. Lett. B",
    volume = "838",
    pages = "137710",
    year = "2023"
}

@article{Gadbail:2023mvu,
    author = "Gadbail, Gaurav N. and De, Avik and Sahoo, P. K.",
    title = "{Cosmological reconstruction and $\Lambda $CDM universe in $f(Q,\,C)$ gravity}",
    eprint = "2312.02492",
    archivePrefix = "arXiv",
    primaryClass = "gr-qc",
    doi = "10.1140/epjc/s10052-023-12288-y",
    journal = "Eur. Phys. J. C",
    volume = "83",
    number = "12",
    pages = "1099",
    year = "2023"
}

@article{Kaczmarek:2024yju,
    author = "Kaczmarek, Adam Z.",
    title = "{Mimetic-f(Q) gravity: Cosmic reconstruction and energy conditions}",
    eprint = "2401.04084",
    archivePrefix = "arXiv",
    primaryClass = "gr-qc",
    doi = "10.1016/j.nuclphysb.2024.116677",
    journal = "Nucl. Phys. B",
    volume = "1007",
    pages = "116677",
    year = "2024"
}

@article{Hu:2022anq,
    author = "Hu, Kun and Katsuragawa, Taishi and Qiu, Taotao",
    title = "{ADM formulation and Hamiltonian analysis of f(Q) gravity}",
    eprint = "2204.12826",
    archivePrefix = "arXiv",
    primaryClass = "gr-qc",
    doi = "10.1103/PhysRevD.106.044025",
    journal = "Phys. Rev. D",
    volume = "106",
    number = "4",
    pages = "044025",
    year = "2022"
}

@article{Nojiri:2010wj,
    author = "Nojiri, Shin'ichi and Odintsov, Sergei D.",
    title = "{Unified cosmic history in modified gravity: from F(R) theory to Lorentz non-invariant models}",
    eprint = "1011.0544",
    archivePrefix = "arXiv",
    primaryClass = "gr-qc",
    doi = "10.1016/j.physrep.2011.04.001",
    journal = "Phys. Rept.",
    volume = "505",
    pages = "59--144",
    year = "2011"
}

@article{Capozziello:2011et,
    author = "Capozziello, Salvatore and De Laurentis, Mariafelicia",
    title = "{Extended Theories of Gravity}",
    eprint = "1108.6266",
    archivePrefix = "arXiv",
    primaryClass = "gr-qc",
    doi = "10.1016/j.physrep.2011.09.003",
    journal = "Phys. Rept.",
    volume = "509",
    pages = "167--321",
    year = "2011"
}

@article{Nojiri:2017ncd,
    author = "Nojiri, S. and Odintsov, S. D. and Oikonomou, V. K.",
    title = "{Modified Gravity Theories on a Nutshell: Inflation, Bounce and Late-time Evolution}",
    eprint = "1705.11098",
    archivePrefix = "arXiv",
    primaryClass = "gr-qc",
    reportNumber = "PHYS.REPT.-692-(2017)-1-104, Phys.Rept. 692 (2017) 1-104",
    doi = "10.1016/j.physrep.2017.06.001",
    journal = "Phys. Rept.",
    volume = "692",
    pages = "1--104",
    year = "2017"
}

@article{Padmanabhan:2009vy,
    author = "Padmanabhan, T.",
    title = "{Thermodynamical Aspects of Gravity: New insights}",
    eprint = "0911.5004",
    archivePrefix = "arXiv",
    primaryClass = "gr-qc",
    doi = "10.1088/0034-4885/73/4/046901",
    journal = "Rept. Prog. Phys.",
    volume = "73",
    pages = "046901",
    year = "2010"
}

@article{Kallosh:2025rni,
    author = "Kallosh, Renata and Linde, Andrei and Roest, Diederik",
    title = "{Atacama Cosmology Telescope, South Pole Telescope, and Chaotic Inflation}",
    eprint = "2503.21030",
    archivePrefix = "arXiv",
    primaryClass = "hep-th",
    doi = "10.1103/d6gn-78hn",
    journal = "Phys. Rev. Lett.",
    volume = "135",
    number = "16",
    pages = "161001",
    year = "2025"
}

@article{Gao:2025onc,
    author = "Gao, Qing and Gong, Yungui and Yi, Zhu and Zhang, Fengge",
    title = "{Nonminimal coupling in light of ACT data}",
    eprint = "2504.15218",
    archivePrefix = "arXiv",
    primaryClass = "astro-ph.CO",
    doi = "10.1016/j.dark.2025.102106",
    journal = "Phys. Dark Univ.",
    volume = "50",
    pages = "102106",
    year = "2025"
}

@article{Liu:2025qca,
    author = "Liu, Lang and Yi, Zhu and Gong, Yungui",
    title = "{Reconciling Nonminimally Coupled Higgs Inflation with ACT DR6 Observations through Reheating}",
    eprint = "2505.02407",
    archivePrefix = "arXiv",
    primaryClass = "astro-ph.CO",
    month = "5",
    year = "2025"
}

@article{Yogesh:2025wak,
    author = "Yogesh and Mohammadi, Abolhassan and Wu, Qiang and Zhu, Tao",
    title = "{Starobinsky like inflation and EGB Gravity in the light of ACT}",
    eprint = "2505.05363",
    archivePrefix = "arXiv",
    primaryClass = "astro-ph.CO",
    doi = "10.1088/1475-7516/2025/10/010",
    journal = "JCAP",
    volume = "10",
    pages = "010",
    year = "2025"
}

@article{Peng:2025bws,
    author = "Peng, Zhi-Zhang and Chen, Zu-Cheng and Liu, Lang",
    title = "{Polynomial potential inflation in the ACT era: From CMB to primordial black holes}",
    eprint = "2505.12816",
    archivePrefix = "arXiv",
    primaryClass = "astro-ph.CO",
    doi = "10.1103/hzcf-q2rk",
    journal = "Phys. Rev. D",
    volume = "113",
    number = "6",
    pages = "063527",
    year = "2026"
}

@article{Yin:2025rrs,
    author = "Yin, Wen",
    title = "{Higgs-like inflation ACTivated mass}",
    eprint = "2505.03004",
    archivePrefix = "arXiv",
    primaryClass = "hep-ph",
    doi = "10.1088/1475-7516/2025/09/062",
    journal = "JCAP",
    volume = "09",
    pages = "062",
    year = "2025"
}

@article{Byrnes:2025kit,
    author = "Byrnes, Christian T. and Cort{\^e}s, Marina and Liddle, Andrew R.",
    title = "{Curvaton in light of ACT results}",
    eprint = "2505.09682",
    archivePrefix = "arXiv",
    primaryClass = "astro-ph.CO",
    doi = "10.1103/x43p-zm85",
    journal = "Phys. Rev. D",
    volume = "113",
    number = "6",
    pages = "063568",
    year = "2026"
}

@article{Wolf:2025ecy,
    author = "Wolf, William J.",
    title = "{Inflationary attractors and radiative corrections in light of ACT data}",
    eprint = "2506.12436",
    archivePrefix = "arXiv",
    primaryClass = "astro-ph.CO",
    doi = "10.1088/1475-7516/2026/02/088",
    journal = "JCAP",
    volume = "02",
    pages = "088",
    year = "2026"
}

@article{Aoki:2025wld,
    author = "Aoki, Shuntaro and Otsuka, Hajime and Yanagita, Ryota",
    title = "{Higgs-modular inflation}",
    eprint = "2504.01622",
    archivePrefix = "arXiv",
    primaryClass = "hep-ph",
    reportNumber = "RIKEN-iTHEMS-Report-25, KYUSHU-HET-317",
    doi = "10.1103/v4z9-676d",
    journal = "Phys. Rev. D",
    volume = "112",
    number = "4",
    pages = "043505",
    year = "2025"
}

@article{Gao:2025viy,
    author = "Gao, Qing and Qian, Yanjiang and Gong, Yungui and Yi, Zhu",
    title = "{Observational constraints on inflationary models with non-minimally derivative coupling by ACT}",
    eprint = "2506.18456",
    archivePrefix = "arXiv",
    primaryClass = "gr-qc",
    doi = "10.1088/1475-7516/2025/08/083",
    journal = "JCAP",
    volume = "08",
    pages = "083",
    year = "2025"
}

@article{Zahoor:2025nuq,
    author = "Zahoor, Mehnaz and Khan, Suhail and Bhat, Imtiyaz Ahmad",
    title = "{Reconciling fractional power potential and EGB gravity in the light of ACT}",
    eprint = "2507.18684",
    archivePrefix = "arXiv",
    primaryClass = "astro-ph.CO",
    doi = "10.1016/j.jheap.2025.100458",
    journal = "JHEAp",
    volume = "49",
    pages = "100458",
    year = "2026"
}

@article{Ferreira:2025lrd,
    author = "Ferreira, Elisa G. M. and McDonough, Evan and Balkenhol, Lennart and Kallosh, Renata and Knox, Lloyd and Linde, Andrei",
    title = "{BAO-CMB tension and implications for inflation}",
    eprint = "2507.12459",
    archivePrefix = "arXiv",
    primaryClass = "astro-ph.CO",
    doi = "10.1103/lq71-b84v",
    journal = "Phys. Rev. D",
    volume = "113",
    number = "4",
    pages = "043524",
    year = "2026"
}

@article{Mohammadi:2025gbu,
    author = "Mohammadi, Abolhassan and Yogesh and Wang, Anzhong",
    title = "{Power law plateau inflation and primordial gravitational waves in the light of ACT}",
    eprint = "2507.06544",
    archivePrefix = "arXiv",
    primaryClass = "astro-ph.CO",
    doi = "10.1016/j.physletb.2025.140054",
    journal = "Phys. Lett. B",
    volume = "872",
    pages = "140054",
    year = "2026"
}

@article{Choudhury:2025vso,
    author = "Choudhury, Sayantan and Bauyrzhan, Gulnur and Singh, Swapnil Kumar and Yerzhanov, Koblandy",
    title = "{What new physics can we extract from inflation using the ACT DR6 and DESI DR2 Observations?}",
    eprint = "2506.15407",
    archivePrefix = "arXiv",
    primaryClass = "astro-ph.CO",
    doi = "10.1016/j.jheap.2026.100656",
    journal = "JHEAp",
    volume = "54",
    pages = "100656",
    year = "2026"
}

@article{Odintsov:2025wai,
    author = "Odintsov, S. D. and Oikonomou, V. K.",
    title = "{GW170817 Viable Einstein-Gauss-Bonnet inflation compatible with the atacama cosmology telescope data}",
    eprint = "2506.08193",
    archivePrefix = "arXiv",
    primaryClass = "gr-qc",
    doi = "10.1016/j.physletb.2025.139779",
    journal = "Phys. Lett. B",
    volume = "868",
    pages = "139779",
    year = "2025"
}

@article{Q:2025ycf,
    author = "Q., Roberto D. Alba and Chagoya, Javier and Roque, Armando A.",
    title = "{Compact stars in Einstein-scalar-Gauss-Bonnet gravity: Regular and divergent scalar field configurations}",
    eprint = "2508.13273",
    archivePrefix = "arXiv",
    primaryClass = "gr-qc",
    doi = "10.1103/ts7l-tr7m",
    journal = "Phys. Rev. D",
    volume = "112",
    number = "8",
    pages = "084029",
    year = "2025"
}

@article{Kouniatalis:2025orn,
    author = "Kouniatalis, Gerasimos and Saridakis, Emmanuel N.",
    title = "{Inflation from a generalized exponential plateau: towards extra suppressed tensor-to-scalar ratios}",
    eprint = "2507.17721",
    archivePrefix = "arXiv",
    primaryClass = "astro-ph.CO",
    doi = "10.1088/1475-7516/2025/11/038",
    journal = "JCAP",
    volume = "11",
    pages = "038",
    year = "2025"
}

@article{Hai:2025wvs,
    author = "Hai, Mishaal and Kamal, Ahmed Rakin and Shamma, Noshin Ferdous and Shuvo, Md Shaikot Jahan",
    title = {{Perturbative K{\"a}hler Moduli Inflation}},
    eprint = "2506.08083",
    archivePrefix = "arXiv",
    primaryClass = "hep-th",
    month = "6",
    year = "2025"
}

@article{Dioguardi:2025vci,
    author = "Dioguardi, Christian and Iovino, Antonio J. and Racioppi, Antonio",
    title = "{Fractional attractors in light of the latest ACT observations}",
    eprint = "2504.02809",
    archivePrefix = "arXiv",
    primaryClass = "gr-qc",
    doi = "10.1016/j.physletb.2025.139664",
    journal = "Phys. Lett. B",
    volume = "868",
    pages = "139664",
    year = "2025"
}

@article{Yuennan:2025kde,
    author = "Yuennan, Jureeporn and Koad, Peeravit and Atamurotov, Farruh and Channuie, Phongpichit",
    title = "{Quantum-corrected $\phi ^{4}$ inflation in light of ACT observations}",
    eprint = "2508.17263",
    archivePrefix = "arXiv",
    primaryClass = "astro-ph.CO",
    doi = "10.1140/epjc/s10052-025-15060-6",
    journal = "Eur. Phys. J. C",
    volume = "85",
    number = "11",
    pages = "1307",
    year = "2025"
}

@article{Ajith:2025rvf,
    author = "Ajith, Abhijith and Kuralkar, Hardik Jitendra and Panda, Sukanta and Vidyarthi, Archit",
    title = "{Eff-ACT-ive Starobinsky pre-inflation}",
    eprint = "2504.15061",
    archivePrefix = "arXiv",
    primaryClass = "gr-qc",
    month = "4",
    year = "2025"
}

@article{Kuralkar:2025hoz,
    author = "Kuralkar, Hardik Jitendra and Panda, Sukanta and Vidyarthi, Archit",
    title = "{Observable primordial gravitational waves from non-minimally coupled R $^{2}$ Palatini modified gravity}",
    eprint = "2502.03573",
    archivePrefix = "arXiv",
    primaryClass = "astro-ph.CO",
    doi = "10.1088/1475-7516/2025/05/073",
    journal = "JCAP",
    volume = "05",
    pages = "073",
    year = "2025"
}

@article{Modak:2025bjv,
    author = "Modak, Tanmoy",
    title = "{R2-Higgs inflation: R3 contribution and preheating after ACT and SPT data}",
    eprint = "2509.02979",
    archivePrefix = "arXiv",
    primaryClass = "hep-ph",
    doi = "10.1103/srpt-jd6s",
    journal = "Phys. Rev. D",
    volume = "112",
    number = "11",
    pages = "115006",
    year = "2025"
}

@article{Aoki:2025ywt,
    author = "Aoki, Shuntaro and Otsuka, Hajime and Yanagita, Ryota",
    title = "{Heavy field effects on inflationary models in light of ACT data}",
    eprint = "2509.06739",
    archivePrefix = "arXiv",
    primaryClass = "hep-ph",
    reportNumber = "RIKEN-iTHEMS-Report-25",
    doi = "10.1088/1475-7516/2025/11/088",
    journal = "JCAP",
    volume = "11",
    pages = "088",
    year = "2025"
}

@article{Ahghari:2025hfy,
    author = "Ahghari, Zahra and Farhoudi, Mehrdad",
    title = "{Inflation with Gauss{\textendash}Bonnet correction and Higgs potential}",
    eprint = "2512.12286",
    archivePrefix = "arXiv",
    primaryClass = "gr-qc",
    doi = "10.1140/epjc/s10052-026-15854-2",
    journal = "Eur. Phys. J. C",
    volume = "86",
    number = "6",
    pages = "601",
    year = "2026"
}

@article{NooriGashti:2025gug,
    author = "Noori Gashti, Saeed and Afshar, Mohammad Ali S. and Alipour, Mohammad Reza and Pourhassan, Behnam and Sadeghi, Jafar",
    title = "{From minimal Higgs inflation with $(R^2)$ term in palatini gravity to Swampland conjectures under ACT constraints}",
    doi = "10.1140/epjc/s10052-025-15066-0",
    journal = "Eur. Phys. J. C",
    volume = "85",
    number = "11",
    pages = "1343",
    year = "2025"
}

@article{Deb:2025gtk,
    author = "Deb, Biswajit and Deshamukhya, Atri",
    title = "{Inflationary models in a minimally coupled $f(R,T)$ gravity: Constraints from $Planck$, BICEP/$Keck$, and ACT}",
    eprint = "2511.06453",
    archivePrefix = "arXiv",
    primaryClass = "gr-qc",
    doi = "10.1016/j.newast.2026.102589",
    journal = "New Astron.",
    volume = "127",
    pages = "102589",
    year = "2026"
}

@article{Ellis:2025zrf,
    author = "Ellis, John and Garcia, Marcos A. G. and Olive, Keith A. and Verner, Sarunas",
    title = "{Constraints on attractor models of inflation and reheating from Planck, BICEP/Keck, ACT DR6, and SPT-3G data}",
    eprint = "2510.18656",
    archivePrefix = "arXiv",
    primaryClass = "hep-ph",
    reportNumber = "UMN-TH-4512/25, FTPI-MINN-25/14, KCL-PH-TH/2025-42, CERN-TH-2025-199",
    doi = "10.1103/d35r-7bn8",
    journal = "Phys. Rev. D",
    volume = "113",
    number = "6",
    pages = "063571",
    year = "2026"
}

@article{Iacconi:2025odq,
    author = "Iacconi, Laura and Bhattacharya, Sukannya and Fasiello, Matteo and Wands, David",
    title = "{Closing in on $α$-attractors}",
    eprint = "2511.14673",
    archivePrefix = "arXiv",
    primaryClass = "astro-ph.CO",
    month = "11",
    year = "2025"
}

@article{Wang:2025dbj,
    author = "Wang, Xinpeng and Kohri, Kazunori and Yanagida, Tsutomu T.",
    title = "{Primordial black holes save R $^{2}$ inflation}",
    eprint = "2506.06797",
    archivePrefix = "arXiv",
    primaryClass = "astro-ph.CO",
    reportNumber = "KEK-TH-2730, KEK-Cosmo-0382",
    doi = "10.1088/1475-7516/2026/03/050",
    journal = "JCAP",
    volume = "03",
    pages = "050",
    year = "2026"
}

@article{Asaka:2015vza,
    author = "Asaka, Takehiko and Iso, Satoshi and Kawai, Hikaru and Kohri, Kazunori and Noumi, Toshifumi and Terada, Takahiro",
    title = "{Reinterpretation of the Starobinsky model}",
    eprint = "1507.04344",
    archivePrefix = "arXiv",
    primaryClass = "hep-th",
    reportNumber = "KEK-TH-1846, RIKEN-QHP-195, RIKEN-STAMP-11, UT-15-23, DESY-15-114, KEK-COSMO-179, COSMO-KOBE-16-13, APCTP-PRE2016-20, KIAS-P16075",
    doi = "10.1093/ptep/ptw161",
    journal = "PTEP",
    volume = "2016",
    number = "12",
    pages = "123E01",
    year = "2016"
}

@article{Oikonomou:2025htz,
    author = "Oikonomou, V. K.",
    title = "{Strong gravity effects on R2-corrected single scalar field inflation and compatibility with the ACT data}",
    eprint = "2508.17363",
    archivePrefix = "arXiv",
    primaryClass = "gr-qc",
    doi = "10.1016/j.physletb.2025.139972",
    journal = "Phys. Lett. B",
    volume = "871",
    pages = "139972",
    year = "2025"
}

@article{Choudhury:2025hnu,
    author = "Choudhury, Sayantan and Singh, Swapnil Kumar and Sahoo, Satish Kumar",
    title = "{Quintessential Inflation in Light of ACT DR6}",
    eprint = "2511.19898",
    archivePrefix = "arXiv",
    primaryClass = "gr-qc",
    month = "11",
    year = "2025"
}

@article{Singh:2025uyr,
    author = "Singh, Swapnil Kumar",
    title = "{Symmetry-Protected $α$-Attractor Hybrid Inflation in Supergravity and Constraints from ACT DR6 and DESI DR2}",
    eprint = "2511.05545",
    archivePrefix = "arXiv",
    primaryClass = "hep-ph",
    month = "10",
    year = "2025"
}

@article{Kim:2025dyi,
    author = "Kim, Jinsu and Wang, Xinpeng and Zhang, Ying-li and Ren, Zhongzhou",
    title = "{Enhancement of primordial curvature perturbations in R $^{3}$-corrected Starobinsky-Higgs inflation}",
    eprint = "2504.12035",
    archivePrefix = "arXiv",
    primaryClass = "astro-ph.CO",
    doi = "10.1088/1475-7516/2025/09/011",
    journal = "JCAP",
    volume = "09",
    pages = "011",
    year = "2025"
}

@article{Garny:2026gcs,
    author = "Garny, Mathias and Niedermann, Florian and Sloth, Martin S.",
    title = "{The End of the First Act: Spectral Running, Interacting Dark Radiation, and the Hubble Tension in Light of ACT DR6 Data}",
    eprint = "2604.26541",
    archivePrefix = "arXiv",
    primaryClass = "astro-ph.CO",
    month = "4",
    year = "2026"
}

@article{Odintsov:2026doe,
    author = "Odintsov, S. D. and Oikonomou, V. K.",
    title = "{R2-corrected Tachyon Scalar Field Inflation, the ACT Data, and Phantom Transition}",
    eprint = "2601.21364",
    archivePrefix = "arXiv",
    primaryClass = "gr-qc",
    doi = "10.1016/j.nuclphysb.2026.117384",
    journal = "Nucl. Phys. B",
    volume = "1025",
    pages = "117384",
    year = "2026"
}

@article{DOnofrio:2025bol,
    author = "D'Onofrio, Simone and Odintsov, Sergei and Paul, Tanmoy",
    title = "{Fitting NANOGrav 15-year data and ACT data with modified inflation in entropic cosmology}",
    eprint = "2510.20484",
    archivePrefix = "arXiv",
    primaryClass = "gr-qc",
    doi = "10.1103/ktt2-lm19",
    journal = "Phys. Rev. D",
    volume = "113",
    number = "4",
    pages = "043527",
    year = "2026"
}

@article{Odintsov:2025jky,
    author = "Odintsov, S. D. and Oikonomou, V. K.",
    title = "{Confronting rainbow-deformed f(R) gravity with the ACT data}",
    eprint = "2508.17358",
    archivePrefix = "arXiv",
    primaryClass = "gr-qc",
    doi = "10.1016/j.physletb.2025.139909",
    journal = "Phys. Lett. B",
    volume = "870",
    pages = "139909",
    year = "2025"
}

@article{Oikonomou:2026qkj,
    author = "Oikonomou, V. K.",
    title = "{Reconstructing ACT-compatible and GW170817-compatible Einstein-Gauss-Bonnet Inflation from the Observational Indices}",
    eprint = "2605.27727",
    archivePrefix = "arXiv",
    primaryClass = "gr-qc",
    doi = "10.1016/j.aop.2026.170547",
    journal = "Annals Phys.",
    volume = "492",
    pages = "170547",
    year = "2026"
}

@article{Odintsov:2026cxz,
    author = "Odintsov, S. D. and Oikonomou, V. K. and Tsyba, Pyotr and Razina, Olga and Rakhatov, Dauren",
    title = "{String-inspired Gauss-Bonnet Gravity Inflation and ACT}",
    eprint = "2604.18861",
    archivePrefix = "arXiv",
    primaryClass = "gr-qc",
    doi = "10.1016/j.dark.2026.102348",
    journal = "Phys. Dark Univ.",
    volume = "52",
    pages = "102348",
    year = "2026"
}

@book{Birrell:1982ix,
    author = "Birrell, N. D. and Davies, P. C. W.",
    title = "{Quantum Fields in Curved Space}",
    doi = "10.1017/CBO9780511622632",
    isbn = "978-0-511-62263-2, 978-0-521-27858-4",
    publisher = "Cambridge University Press",
    address = "Cambridge, UK",
    series = "Cambridge Monographs on Mathematical Physics",
    year = "1982"
}

@article{Parker:1968mv,
    author = "Parker, L.",
    title = "{Particle creation in expanding universes}",
    doi = "10.1103/PhysRevLett.21.562",
    journal = "Phys. Rev. Lett.",
    volume = "21",
    pages = "562--564",
    year = "1968"
}

@article{Parker:1969au,
    author = "Parker, Leonard",
    title = "{Quantized fields and particle creation in expanding universes. 1.}",
    doi = "10.1103/PhysRev.183.1057",
    journal = "Phys. Rev.",
    volume = "183",
    pages = "1057--1068",
    year = "1969"
}

@article{Parker:1971pt,
    author = "Parker, L.",
    title = "{Quantized fields and particle creation in expanding universes. 2.}",
    doi = "10.1103/PhysRevD.3.346",
    journal = "Phys. Rev. D",
    volume = "3",
    pages = "346--356",
    year = "1971",
    note = "[Erratum: Phys.Rev.D 3, 2546--2546 (1971)]"
}

@book{Buchbinder:1992gdx,
    author = "Buchbinder, I. L. and Odintsov, S. D. and Shapiro, I. L.",
    title = "{Effective Action in Quantum Gravity}",
    doi = "10.1201/9780203758922",
    isbn = "978-0-203-75892-2, 9780750301228, 978-0-7503-0122-0",
    publisher = "Routledge",
    month = "9",
    year = "2017"
}

@article{Nojiri:2024hau,
    author = "Nojiri, Shin'ichi and Odintsov, Sergei D.",
    title = "{F(Q) Gravity with Gauss{\textendash}Bonnet Corrections: From Early-Time Inflation to Late-Time Acceleration}",
    eprint = "2406.12558",
    archivePrefix = "arXiv",
    primaryClass = "gr-qc",
    reportNumber = "KEK-TH-2629, KEK-Cosmo-0346",
    doi = "10.1002/prop.202400113",
    journal = "Fortsch. Phys.",
    volume = "72",
    number = "9-10",
    pages = "2400113",
    year = "2024"
}

@article{Capozziello:2023vne,
    author = "Capozziello, Salvatore and De Falco, Vittorio and Ferrara, Carmen",
    title = "{The role of the boundary term in f(Q,~B) symmetric teleparallel gravity}",
    eprint = "2307.13280",
    archivePrefix = "arXiv",
    primaryClass = "gr-qc",
    doi = "10.1140/epjc/s10052-023-12072-y",
    journal = "Eur. Phys. J. C",
    volume = "83",
    number = "10",
    pages = "915",
    year = "2023"
}

@article{Odintsov:2025bmp,
    author = "Odintsov, Sergei D. and Paul, Tanmoy",
    title = "{ACT inflation and its influence on reheating era in Einstein-Gauss-Bonnet gravity}",
    eprint = "2508.11377",
    archivePrefix = "arXiv",
    primaryClass = "gr-qc",
    doi = "10.1016/j.physletb.2025.139930",
    journal = "Phys. Lett. B",
    volume = "870",
    pages = "139930",
    year = "2025"
}

@article{AtacamaCosmologyTelescope:2025vnj,
    author = "Naess, Sigurd and others",
    collaboration = "Atacama Cosmology Telescope",
    title = "{The Atacama Cosmology Telescope: DR6 maps}",
    eprint = "2503.14451",
    archivePrefix = "arXiv",
    primaryClass = "astro-ph.CO",
    reportNumber = "FERMILAB-PUB-25-0160-PPD",
    doi = "10.1088/1475-7516/2025/11/061",
    journal = "JCAP",
    volume = "11",
    pages = "061",
    year = "2025"
}

@article{Sabogal:2026qvy,
    author = "Sabogal, Miguel A. and Iovino, Antonio J. and Vagnozzi, Sunny",
    title = "{Running into tension: primordial black holes from ultra-slow-roll inflation, spectral running, and the Hubble tension}",
    eprint = "2606.31362",
    archivePrefix = "arXiv",
    primaryClass = "astro-ph.CO",
    month = "6",
    year = "2026"
}

@article{CosmoVerseNetwork:2025alb,
    author = "Di Valentino, Eleonora and others",
    collaboration = "CosmoVerse Network",
    title = "{The CosmoVerse White Paper: Addressing observational tensions in cosmology with systematics and fundamental physics}",
    eprint = "2504.01669",
    archivePrefix = "arXiv",
    primaryClass = "astro-ph.CO",
    doi = "10.1016/j.dark.2025.101965",
    journal = "Phys. Dark Univ.",
    volume = "49",
    pages = "101965",
    year = "2025"
}

@article{Hwang:1990re,
    author = "Hwang, J. C.",
    title = "{Cosmological perturbations in generalized gravity theories: Formulation}",
    doi = "10.1088/0264-9381/7/9/013",
    journal = "Class. Quant. Grav.",
    volume = "7",
    pages = "1613--1631",
    year = "1990"
}

@article{Hwang:1990jh,
    author = "Hwang, J. C.",
    title = "{Cosmological perturbations in generalized gravity theories: Solutions}",
    doi = "10.1103/PhysRevD.42.2601",
    journal = "Phys. Rev. D",
    volume = "42",
    pages = "2601--2606",
    year = "1990"
}

@article{Makino:1991sg,
    author = "Makino, Nobuyosi and Sasaki, Misao",
    title = "{The Density perturbation in the chaotic inflation with nonminimal coupling}",
    reportNumber = "YITP-U-91-8",
    doi = "10.1143/PTP.86.103",
    journal = "Prog. Theor. Phys.",
    volume = "86",
    pages = "103--118",
    year = "1991"
}

\end{document}